\documentclass[a4paper,11pt]{article}
\pdfoutput=1 % if your are submitting a pdflatex (i.e. if you have
\usepackage{jcappub} % for details on the use of the package, please
\usepackage[T1]{fontenc} % if needed
\usepackage{ulem}

\allowdisplaybreaks[1]

\title{\boldmath On the importance of heavy fields in pseudo-scalar inflation}

\author[a,1]{Chong-Bin Chen,\note{Corresponding author.}}
\author[b,a]{Ziwei Wang}
\author[c,a]{and Siyi Zhou}

\affiliation[a]{Department of Physics, Kobe University, Kobe 657-8501, Japan}
\affiliation[b]{Key Laboratory of Dark Matter and Space Astronomy, Purple Mountain Observatory, Chinese Academy of Sciences, Nanjing 210023,
China}
\affiliation[c]{Department of Physics and Chongqing Key Laboratory for Strongly Coupled Physics, Chongqing University, Chongqing 401331, China}

\emailAdd{chongbin@stu.kobe-u.ac.jp}
\emailAdd{zwwang@pmo.ac.cn}
\emailAdd{siyi@cqu.edu.cn}

\abstract{Pseudo-scalar inflation coupled with U(1) gauge fields through the Chern-Simons term has been extensively studied. However, new physics arising from UV theories may still influence the pseudo-scalar field at low-energy scales, potentially impacting predictions of inflation. In the realm of effective field theory (EFT), we investigated axion inflation, where operators from heavy fields are also present, in addition to the axion and gauge fields.  The integrated out fields have two significant effects: the non-linear dispersion regime and coupling heavy modes to the Chern-Simons term. The first effect changes the propagation of the curvature fluctuation, while the second one results in additional
operators that contribute to curvature fluctuation via inverse decay. We derived the power spectrum and magnitude of equilateral non-Gaussianity in this low-energy EFT. We found that the second effect could become significant as the mass of heavy fields approaches Hubble scale.}

\begin{document}
\maketitle
\flushbottom

\section{Introduction}
\label{sec:intro}
Axion particles are initially proposed to solve the strong CP problem in the QCD physics \cite{Weinberg:1975ui,tHooft:1976rip,tHooft:1976snw,Adler:1969gk,Bell:1969ts,Bardeen:1969md}. In string theory, such axion-like particles are produced by string
compactification to a low-energy effective theory. The mass of axion-like particles is logarithmically distributed over a very wide mass range \cite{Arvanitaki:2009fg}, which gives rise to rich
phenomenology in a wide range of periods in the universe (\cite{Marsh:2015xka,Hui:2016ltb,Niemeyer:2019aqm} are reviews of axion in cosmology). One of the most important case is that axion can be one of the candidates of inflaton in the very early universe \cite{Freese:1990rb,Adams:1992bn,Moroi:2000jr,Dimopoulos:2005ac}, due to protecting naturalness from higher operators by shift-symmetry of the axion. However, the theoretical requirement from string theory $f\lesssim M_{\text{pl}}$ of the original natural inflation is in tension with the Planck data \cite{Planck:2018jri}. Hence, introducing additional dynamics \textemdash the coupling with gauge fields, is studied during the inflationary period \cite{Garretson:1992vt,Anber:2006xt,Durrer:2010mq,Barnaby:2010vf,Barnaby:2011vw,Barnaby:2011qe} and after the end of inflation \cite{Braden:2010wd,Adshead:2015pva,Adshead:2016iae,Cuissa:2018oiw}. The symmetric interaction with U(1)-gauge fields is 
\begin{equation}
    \mathcal{L}_{\text{Chern-Simons}}\sim-\frac{\alpha}{f}\phi F_{\mu\nu}\tilde{F}^{\mu\nu},
\end{equation}
where $f$ is the decay constant of axion $\phi$ and $\alpha$ is a dimensionless parameter. One of the gauge field helicities has instability and can be exponentially produced during inflation \cite{Garretson:1992vt,Anber:2006xt,Durrer:2010mq}. The characteristic parameter of the enhancement is given by $\xi\equiv \alpha\dot{\phi}/(2fH)$. The interesting phenomenology of this effect is quite rich. The large production of one of the gauge field helicities leads to the large equilateral primordial non-Gaussianity \cite{Barnaby:2010vf,Barnaby:2011vw}, chiral gravitational wave \cite{Barnaby:2011qe,Sorbo:2011rz,Cook:2011hg,Anber:2012du,Bartolo:2016ami,Bastero-Gil:2022fme}, productions of primordial black holes \cite{Linde:2012bt,Bugaev:2013fya,Garcia-Bellido:2016dkw,Domcke:2017fix} and large scale magnetic fields \cite{Anber:2006xt,Durrer:2010mq,Adshead:2016iae,Caprini:2014mja,Fujita:2015iga,Patel:2019isj}. The constrain of non-Gaussianity from CMB limit gives a upper bound of the parameter $\xi\lesssim 2.5$ \cite{Planck:2018jri,Planck:2019kim}. Moreover, the bound of the mass spectrum of primordial black holes provides a tighter constrain  $\xi\lesssim 1.5$ \cite{Linde:2012bt,Bugaev:2013fya}.

In this scenario, the electromagnetic(EM) fields as a source of the curvature fluctuation after enhancements due to the instability. Then the correlation of $k$-modes has additional contribution from the EM fields around Hubble scale $k\sim aH$ for $\xi\gtrsim 1$. This is the so-called inverse decay processes \cite{Barnaby:2010vf,Barnaby:2011vw}. This processes relies on the interacting form between curvature fluctuation and EM fields and the propagation way of modes sourced by EM fields. However, in the effective field theory(EFT) formalism of inflation \cite{Cheung:2007st,Weinberg:2008hq,Senatore:2010wk}, it's known that the light-field (in our discussion, is the curvature fluctuation) could still be sensitive to the heavy degree of freedom (in our discussion, is the entropic fluctuation) in which even has been integrated out \cite{Achucarro:2010da,Achucarro:2010jv}. This implies that there maybe new physics around Hubble scales and hence modify the parametrization of the single-field inflationary observations. Then the impacts of EM fields due to the inverse decay process to modes with Hubble-length wavelength are also need to be carefully  addressed. In this paper, we are going to discuss how the new physics around or even lower than Hubble scale affect the predictability of axion inflation.

Specifically, for example, a rapidly turning of trajectory in the field space could dramatically reduce the speed of sound of the light mode \cite{Achucarro:2012yr,Cespedes:2012hu,Achucarro:2012sm}. This would result in a unacceptable low non-unitarity and strong-coupling scale so the EFT description is failed \cite{Baumann:2011su} if we use the formalism of \cite{Cheung:2007st}. To relax this issue, 
the effects of heavy fields on the low-energy EFT of Nambu-Goldstone boson introducing infinity new higher-order operators to modify the theory. The Lagrangian of these non-universal operators is given by \cite{Gwyn:2012mw}
\begin{align}\label{EFT}
    \mathcal{L}_{\text{higher}}\sim\sum_{n=2}\frac{M_n^4}{n!}\left[(1+g^{00})\frac{M^2}{M^2-\tilde{\nabla}^2}\right]^{n-1}(1+g^{00})+\cdots,
\end{align}
where $M$ is the mass of the heavy field. There is a new physics regime where the propagation of the light mode is characterized by a non-linear dispersion relation
\begin{equation}
    \omega(p)\sim p^2
\end{equation}
even when the physical wavelength of modes is around Hubble length. The experience of this non-linear dispersion regime can lift the strong coupling scale to a value above the symmetry breaking scale of NG boson so that the low-energy EFT formalism is weakly coupled \cite{Baumann:2011su,Gwyn:2012mw}. For more lower energy-scale, this dispersion relation is reduced to a linear one $\omega(p)=c_sp$ but with a reduced speed of sound. Then when the EM fields go through the enhancement period in these regimes, the propagation of modes should be modified compared to the single-field case. Moreover, although the EM field is not directly coupled to other scalar fields, it still has interacting with the heavy mode through the axion field $\delta\phi$ in the unitary gauge. In the low-energy of integrating out this mode, leaving us a new interacting of the light mode and the EM field
\begin{equation}
    \mathcal{L}_{\dot{\zeta}AA}\sim \frac{1}{M^2-\nabla^2/a^2}\dot{\zeta}F_{\mu\nu}\tilde{F}^{\mu\nu},
\end{equation}
where $\zeta$ is the curvature fluctuation. We address the impacts of new vertexes from this interaction for linear and non-linear dispersion regimes. In this paper, we calculate the total power spectrum and the magnitude of equilateral non-Gaussianity of the inverse decay process, due to the peak of the non-Gaussiantiy is the equilateral configuration.

The organization of the paper is as follows. In section \ref{sec:actf}, we review the scenario of axion inflation and the mechanism of particle production of the EM field. In section \ref{sec:ehm}, we study how the heavy modes modify the physics at low-energy scale. In section \ref{sec:cf}, we calculate the two-point and three-point correlation function in the regimes of new physics. In section \ref{sec:ip}, we discuss the impact of the previous results on phenomenology. The final section is devoted to the conclusions.

\section{Axial coupling to the inflation}
\label{sec:actf}
We also consider the EM fields during inflation, which are conformally coupled with gravity hence produced particles decay rapidly. However, the axions also have the Chern-Simons couplings with the $U(1)$ fields. The action of the EM parts are given by
\begin{equation}
    S_{\text{EM}}=\int d^4x\sqrt{-g}\left(-\frac{1}{4}F_{\mu\nu}F^{\mu\nu}-\frac{\alpha}{f}\phi F_{\mu\nu}\tilde{F}^{\mu\nu}\right),
\end{equation}
where $f$ is the decay constant of axion $\phi$ and $\alpha$ is a dimensionless parameter which is model dependent and we generally expect that it has order unity magnitude. The field strengths
\begin{equation}
    F_{\mu\nu}\equiv \partial_{\mu}A_{\nu}-\partial_{\nu}A_{\mu},\ \ \ \ \ \ \ \tilde{F}^{\mu\nu}\equiv \frac{1}{2}\eta^{\mu\nu\alpha\beta}F_{\alpha\beta},
\end{equation}
where $\eta^{\mu\nu\alpha\beta}=\epsilon^{\mu\nu\alpha\beta}/(2\sqrt{-g})$ is the totally anti-symmetric tensor in four dimensional spacetime and $\epsilon^{0123}=1$. Physically, the electric and magnetic field measured by an observer with four-velocity $u^{\mu}=(1,\boldsymbol{0})+\mathcal{O}(1)$ is covariantly defined by 
\begin{equation}
    E_{\mu}\equiv F_{\mu\nu}u^{\nu},\ \ \ \ \ \ B_{\mu}\equiv \frac{1}{2}\eta_{\mu\alpha\beta\nu}F^{\alpha\beta}u^{\nu}.
\end{equation}
We work in Coulomb gauge $A_0$=0 and $\partial_iA_i=0$ and then define the Euclidean three-vector fields $\boldsymbol{E}$ and $\boldsymbol{B}$ through $E_{\mu}=a(0,\boldsymbol{E})$ and $B_{\mu}=a(0,\boldsymbol{B})$.

We study the evolution of the inflation in a flat FRW background
\begin{equation}
    ds^2=-dt^2+a^2(t)\delta_{ij}dx^i dx^j,
\end{equation}
where $a$ is the scale factor and $t$ is the cosmological time. Under this configuration, the constrains equation of spacetime is
\begin{align}\label{ceom}
     3M_{\text{pl}}^2H^2=\frac{1}{2}\left(\dot{\phi}^2+\frac{(\nabla\phi)^2}{a^2}\right)+V(\phi)+\frac{1}{2}\left(\boldsymbol{E}^2+\boldsymbol{B}^2\right)+\cdots,
\end{align}
where ``$\cdots$'' stands for the contributions of heavy degree of freedoms. We here retain the spatial terms of scalar fields because the EM fields are inhomogeneous. The equations of motion of the scalar fields are given by
\begin{align}
    \ddot{\phi}+3H\dot{\phi}-\frac{\nabla^2}{a^2}\phi+V_{\phi}-\frac{\alpha}{f}\boldsymbol{E}\cdot\boldsymbol{B}+\cdots=0~.\label{phieom}
\end{align}
To protect the homogeneous slow-roll inflation, the contributions of background EM fields should be negligible compared to the energy of inflation (see discussions in the section \ref{appendix:be}). Then the EM fields are at least second order quantities.

\subsection{Particle production of electromagnetism}
\label{subsec:ppef}
We now treat the evolution of the EM fields. Although the EM fields are conformally coupled to gravity during inflation, we have also the Chern-Simons coupling to the inflaton and this term can cause instability in one of the modes of EM fields. To see this, we first obtain the equation of motion of EM fields from the perturbed action of $\boldsymbol{A}$,
\begin{equation}\label{EMeom}
    \boldsymbol{A}''-\nabla^2\boldsymbol{A}-\frac{\alpha}{f}\phi'\nabla\times\boldsymbol{A}=0,
\end{equation}
where the prime denotes the derivative of conformal time $\tau=\int{dt/a}$. The third term comes from the coupling with inflaton. We quantize the EM fields as
\begin{equation}\label{qA}
    \boldsymbol{A}(\tau,\boldsymbol{x})=\sum_{\lambda=\pm}\int\frac{d^3k}{(2\pi)^{3}}\left[\boldsymbol{\epsilon}_{\lambda}(\boldsymbol{k})a_{\lambda}(\boldsymbol{k})A_{\lambda}(\tau,\boldsymbol{k})e^{i\boldsymbol{k}\cdot\boldsymbol{x}}+\text{h.c.}\right],
\end{equation}
where the annihilation and creation operator satisfy $[a_{\lambda}(\boldsymbol{k}),a^{\dagger}_{\lambda'}(\boldsymbol{k'})]=(2\pi)^3\delta_{\lambda\lambda'}\delta^3(\boldsymbol{k}-\boldsymbol{k}')$. $\boldsymbol{\epsilon}_\lambda$ are the polarization vectors and have $\boldsymbol{k}\cdot\boldsymbol{\epsilon}_{\lambda}=0$, $\boldsymbol{k}\times\boldsymbol{\epsilon}_{\pm}=\mp i\boldsymbol{\epsilon}_{\pm}$ and $\epsilon_{\lambda}(-\boldsymbol{k})=\epsilon_{\lambda}(\boldsymbol{k})^*$. The EM fields have two polarization modes $A_{\pm}$ and we can insert them in the (\ref{EMeom}) and write down their evolution equations as
\begin{equation}\label{Aeom}
    A_{\pm}''(\tau,k)+\left(k^2\pm\frac{2k\xi}{\tau}\right)A_{\pm}(\tau,k)=0,\ \ \ \ \ \xi\equiv\frac{\alpha\dot{\phi}}{2fH}.
\end{equation}
We can see for the inflation with $\dot{\phi}>0$, as we consider in this paper, the $A_{+}$ mode experiences a tachyonic instability for $-k\tau\lesssim 2\xi$, while the $A_{-}$ mode doesn't. Hence the EM particles are produced during inflation if the inflaton is slowly rolling and we can treat $\xi$ as a constant. 

With the adiabatic initial conditions $A_{+}=e^{-ik\tau}/{\sqrt{2k}}$ on sub-horizon limit $k\tau\rightarrow-\infty$, the solution of (\ref{Aeom}) can approximately described by \cite{Anber:2006xt,Durrer:2010mq}
\begin{align}\label{Asol}
    A_{+}(\tau,k)\simeq \frac{1}{\sqrt{2k}}\left(\frac{-k\tau}{2\xi}\right)^{1/4}e^{\pi\xi-2\sqrt{-2\xi k\tau}}~.
\end{align}
in the interval $(8\xi)^{-1}\lesssim-k\tau\lesssim 2\xi$. The solution has a growing factor $e^{\pi\xi}$. The case $\xi\lesssim1$ is uninteresting hence can be ignored. We can see the mode functions of EM fields are real-valued. The $A_i$ and its conjugate momentum are hence commuting \cite{Barnaby:2011qe},
\begin{equation}\label{commu}
    \left[A_i'(\tau,\boldsymbol{x}),A_j(\tau,\boldsymbol{x}')\right]\simeq0.
\end{equation}
This relations is a good approximation when calculate the correlation function because the modes with $-k\tau>2\xi$ remain in their vacuum hence is not concerted. Only $-k\tau<2\xi$ modes are interesting and they are all commuting variables. One can use the exact solution of (\ref{Aeom}) to perform an accurate calculations for small $\xi$ case \cite{Ballardini:2019rqh,Domcke:2019qmm}. However, we will still use (\ref{Asol}) in this paper because it's accurate enough if effects of heavy fields are important. We should also notice the backreaction of EM fields to the inflaton. To guarantee the sub-domination of the energy density of EM fields compared to the inflaton, the parameter $\xi$ has upper bound to avoid generating too many particles. This bound is given by appendix \ref{appendix:be}. From now on we ignore the subscript ``+'' in $\boldsymbol{\epsilon}_+$ and $A_+$.

\section{Effects of the heavy modes}
\label{sec:ehm}
\subsection{EFT of a single light-field}

We study a specific example of EFT (\ref{EFT}) by integrating out one single heavy field\footnote{The small sound speed can be also achieved without any heavy fields \cite{Jazayeri:2022kjy,Jazayeri:2023xcj}.}. This heavy field is from, for example, deviation of the trajectory of two scalar fields in field space \cite{Achucarro:2010da,Achucarro:2010jv,Achucarro:2012yr,Cespedes:2012hu,Achucarro:2012sm}, which is characterized by the entropic fluctuation $\mathcal{F}$. Starting from the two scalar fields, one can decompose the fluctuations of scalar fields into $\delta\phi^a=N^a\mathcal{F}-\dot{\phi}^a\pi$, where $\mathcal{F}$ is the entropic field and $\pi$ is the NG boson. $N^a$ is the norm vector of the trajectory. Then the Mukhanov-Sasaki variables are defined by
\begin{equation}
    Q^a\equiv N^a\mathcal{F}-\dot{\phi}^a\pi-\frac{\dot{\phi}^a}{H}\zeta.
\end{equation}
Since we ignored the backreaction of the gauge fields, the adiabatic mode is nearly given by two scalar fluctuations. Then the unitary gauge(or comoving gauge) is chosen as $\pi=0$. In this gauge, the quadratic Lagrangian of the heavy field in the decoupling limit is given by \cite{Achucarro:2012yr,Cespedes:2012hu,Achucarro:2012sm}
\begin{equation}
    \mathcal{L}_{\mathcal{F}\mathcal{F}}=\frac{1}{2}a^3\left[\dot{\mathcal{F}}^2-\frac{(\nabla \mathcal{F})^2}{a^2}-M^2\mathcal{F}^2+4\dot{\sigma}\eta_{\bot}\dot{\zeta}\mathcal{F}\right],
\end{equation}
where $\dot{\sigma}^2\equiv G_{ab}\phi^a\phi^b$ and $\eta_{\bot}$ is the turning-rate of trajectory defined by $\eta_{\bot}\equiv -V_N/(H\dot{\sigma})$. From this Lagrangian we can obtain the equation of motion
\begin{equation}
    \left(-\square+M^2\right)\mathcal{F}=2\dot{\sigma}\eta_{\bot}\dot{\zeta}, \ \ \ \ \ \ \ \ \square\equiv-\frac{\partial^2}{\partial t^2}-3H\frac{\partial}{\partial t}+\frac{\nabla^2}{a^2}.
\end{equation}

We are interested in the low-energy EFT where the $\mathcal{F}$ can be integrated out\footnote{The actual integrated out d.o.f are the high-frequency modes in $\zeta$ and $\mathcal{F}$ \cite{Achucarro:2012yr}}. Hence the adiabatic approximation $|\dot{X}/X|\ll |M|$, where $X=\{H,\dot{\sigma},\eta_{\bot} \}$ is assumed. We are in the low-energy $\omega^2\ll M^2+p^2$ where hierarchy of scales between frequency and momentum
\begin{equation}\label{expansion}
    \frac{1}{M^2-\square}=\frac{1}{M^2-\nabla^2/a^2}+\frac{-\partial_t^2-3H\partial_t}{\left(M^2-\nabla^2/a^2\right)^2}+\cdots
\end{equation}
is presented. The ``$\cdots$'' is the higher-order time derivative terms and are suppressed by $\omega/M$. In this regime the equation of motion at leading-order is solved as
\begin{align}\label{LOeom}
    \mathcal{F}=\frac{2\dot{\sigma}\eta_{\bot}\dot{\zeta}}{M^2-\nabla^2/a^2}+\cdots.
\end{align}
Then the quadratic parts of the light field $\zeta$ is reduced to \cite{Gwyn:2012mw}
\begin{equation}\label{EFTUV}
    \mathcal{L}_{\zeta\zeta}=a^3M_{\text{pl}}^2\epsilon\left[\dot{\zeta}\left(\frac{M^2c_s^{-2}-\nabla^2/a^2}{M^2-\nabla^2/a^2}\right)\dot{\zeta}-\frac{(\nabla \mathcal{\zeta})^2}{a^2}\right],
\end{equation}
where $c_s$ is the speed of sound of the fluctuation
\begin{equation}\label{cs}
    \frac{1}{c_s^2}\equiv1+\frac{4H^2\eta_{\bot}^2}{M^2}.
\end{equation}
The sound spped is the function of $\eta_{\bot}$ and entropic mass. It reduces to $c_s=1$ for single-field inflation $\eta_{\bot}=0$. 

The coupling with heavy fields change the dispersion relation of the propagation of the light field, which should be written as
\begin{equation}\label{dispersion}
    \omega^2(p)=\frac{M^2+p^2}{M^2c_s^{-2}+p^2}p^2,
\end{equation}
where $p=k/a$ is the physical momentum. The low-energy limit $\omega^2\ll M^2+p^2$ introduces a UV cut-off on momentum and   energy scale
\begin{equation}
    p_{\text{UV}}\sim Mc_s^{-1},\ \ \ \ \ \ \Lambda_{\text{UV}}\sim\omega(p_{\text{UV}}^2)\sim Mc_s^{-1}.
\end{equation}
Above this scale, the expansion (\ref{expansion}) is failed and the completed UV theory should be taken into account. Below this scale, although the heavy mode has been integrated out, the dispersion relation is generally still non-linear 
\begin{equation}\label{nld}
    \omega^2(p)=c_s^2p^2+\frac{1-c_s^2}{M^2c_s^{-2}}p^4+\mathcal{O}(p^6)
\end{equation}
The domination of non-linear term introduces another momentum and energy scale
\begin{equation}\label{linearEFT}
     p_{\text{new}}\sim M,\ \ \ \ \ \ \Lambda_{\text{new}}\sim \omega({p_{\text{new}}^2})\sim Mc_s.
\end{equation}
Above this scale the non-linear dispersion relation due to the heavy modes become important so the propagation of the light fields is quite different from the single-field inflation. The speed of sound $c_s$ determines such kind of effects from heavy fields and also the hierarchy between $\Lambda_{\text{UV}}$ and $\Lambda_{\text{new}}$ (also between $p_{\text{UV}}$ and $p_{\text{new}}$). Below scales $\Lambda_{\text{new}}$ the momentum in dispersion relation is also unimportant compared to $M$. Then the effective field theory of $\zeta$ is reduced to
\begin{equation}\label{zeta2}
    \mathcal{L}_{\zeta\zeta}\simeq a^3M_{\text{pl}}^2\frac{\epsilon}{c_s^2}\left(\dot{\zeta}^2-c_s^2\frac{\left(\nabla\zeta\right)^2}{a^2}\right), \ \ \ \ \ \ \left(\omega\ll \Lambda_{\text{new}}\right)
\end{equation}
with linear dispersion relation $\omega^2=c_s^2p^2$. We obtained a single-field theory with a reduced speed of sound. The speed of sound exposed many important features of the UV theory hence we should carefully compare it with the parameters of the coupled gauge field. The effect from gauge fields to the power spectrum is sensitive to the propagation of the light field.

We should carefully discuss the valid of the EFT description when the EM fields matter.
\begin{itemize}
\item[(1)] Linear dispersion regime: $p_{\text{new}}\gg 2\xi H$. In this regime when the particle production of EM fields is sufficiently large, the EFT of NG boson has linear dispersion relation. Then we can discuss the effects from EM field to the observations in the EFT (\ref{zeta2}) with a reduced sound speed $c_s<1$. 

\item[(2)] Non-linear dispersion regime: $p_{\text{UV}}\gg 2\xi H\gg p_{\text{new}}$. In this regime the non-linear dispersion relation from the heavy field becomes important and the way that EM field impacts observations should be described by EFT (\ref{EFTUV}). 

\item[(3)] UV-completed regime: $2\xi H\gg p_{\text{UV}}$. In this regime we need the UV theory of the EFT and concert the specific interaction of heavy fields with the EM field.

\end{itemize}

\subsection{Inverse decay processes}
Now we threat the coupling of metric field with EM fields. We are only considering the situation up to non-Gaussianity, where the lapse function $N=1+A$ and the shift vector $N^i=\delta^{ij}\partial_{j}B/a^2$ only include first order contribution. Hence we don't need to consider the contributions from the EM fields, which are at least second order contributions to the constrained systems. In the comoving gauge we have 
\begin{equation}\label{constrains}
    A=\frac{\dot{\zeta}}{H},\ \ \ \ \ B=-\frac{\zeta}{H}+\theta,\ \ \ \ \ \frac{\partial^2\theta}{a^2}=\epsilon\dot{\zeta}+\frac{\dot{\sigma}\eta_{\bot}}{M_{\text{pl}}^2}\mathcal{F}.
\end{equation}
Moreover, because we have ignored the backreaction  of EM fields to the inflation, then the quadratic and cubic action are the same as the only multi-scalar system. The new contributions are from the interactions of EM fields with curvature and entropic fluctuations, as we will see later. 
%The ADM formalism of the $U(1)$ gauge field Lagrangian is given by
%\begin{align}\label{EMLa}
%    \mathcal{L}_{\text{EM}}=&\frac{\sqrt{g}}{2N}g^{ik}(E_i+F_{ij}N^j)(E_k+F_{kl}N^l)-\frac{N\sqrt{g}}{4}g^{ik}g^{jl}F_{ij}F_{kl}\nonumber\\
%    &-\frac{\alpha \phi}{f_a}\eta^{\mu\nu\alpha\beta}F_{\mu\nu}F_{\alpha\beta}.
%\end{align}
Firstly we have quadratic Lagrangian of EM fields
\begin{equation}
    \mathcal{L}_{AA}=\frac{a}{2}\dot{A_i}\dot{A_i}-\frac{1}{2a}\partial_{i}A_{j}\partial_{i}A_{j}+\frac{\alpha}{f_{a}}\dot{\phi}\epsilon^{ijk}A_{i}\partial_{j}A_{k},
\end{equation}
where we have dropped out the boundary terms. The linear equations of motion (\ref{EMeom}) is obtained by varying this quadratic action.

Then we consider the cubic Lagrangian. First is the interactions of EM fields and the metric fluctuation
\begin{align}\label{gAA}
    \mathcal{L}_{\zeta AA}=&\frac{\zeta}{H}\frac{\alpha}{f_a}\dot{\phi}\epsilon^{ijk}\dot{A_i}\partial_jA_k+\frac{a}{H}\zeta\dot{A_i}\frac{\delta\mathcal{L}^{(2)}_{AA}}{\delta A_i}+\partial_tD\nonumber\\
    &+\frac{1}{a}\dot{A_i}\left(\partial_{i}A_j-\partial_{j}A_i\right)\partial_{j}\theta+\epsilon\left[\frac{\zeta}{2a}\left(a^2\dot{A_i}\dot{A_i}+\partial_{i}A_{j}\partial_{i}A_{j}-\partial_{i}A_{j}\partial_{j}A_{i}\right)\right],
\end{align}
where we have inserted constrains $A$ and $B$ (\ref{constrains}) and ignored the spatial boundary terms because they do not contribute to the correlation functions. The final term $\mathcal{O}(\epsilon)$ are the terms with factors of slow-roll parameter $\epsilon$ hence their contribution is subleading. One can use the linear equation of motion to eliminate the third term, but we should keep the boundary term of total time derivative, just like the curvature fluctuation in the single-field inflation \cite{Maldacena:2002vr,Arroja:2011yj,Renaux-Petel:2011zgy}. The temporal boundary term is 
\begin{equation}
    D\equiv-\frac{a\zeta}{2H}\dot{A_i}\dot{A_i}+\frac{\zeta}{2aH}\left(\partial_{i}A_{j}\partial_{i}A_{j}-\partial_{i}A_{j}\partial_{j}A_{i}\right)\propto{a^3}(\boldsymbol{E}^2+\boldsymbol{B}^2).
\end{equation}
From equation (\ref{EMeom}), one can find that the solutions $A_i$ become constants on super-horizon limit, which means the physical EM fields decay sufficiently outside the horizon \cite{Barnaby:2011vw}. Hence when we calculate the correlation functions at late time, this boundary term has vanishing contribution. 

We are using the comoving gauge $\pi=0$ of two-scalar system and only consider the axial coupling of inflaton to the EM fields hence only $\delta\phi$ is included. This gauge implies the relation between fluctuations of two scalar fields $\dot{\phi}_a\delta\phi^a=0$. Without losing generality, for field space $G_{ab}=\text{diag}\{1,K^2\}$ and the second scalar field is $\chi$, the entropic fluctuation can be represented as
\begin{equation}\label{Fphi}
    \mathcal{F}= \frac{\dot{\sigma}}{h\dot{\phi}}\delta\phi,\ \ \ \ \ \ \ h\equiv-\frac{K\dot{\chi}}{\dot{\phi}}.
\end{equation}
Then we can replace the inflaton fluctuation $\delta\phi$ in Lagrangian by the entropic fluctuation $\mathcal{F}$ 
\begin{align}
    \mathcal{L}_{\mathcal{F}AA}=\frac{h\dot{\phi}}{\dot{\sigma}}\frac{\alpha}{f}\mathcal{F}\epsilon^{ijk}\dot{A_i}\partial_{j}A_k.
\end{align}
If we consider the heavy entropic fluctuation, the $\mathcal{F}$ can be integrated out by using the leading order equation of motion (\ref{LOeom}). Then this term becomes the interaction of EM fields and time derivative of $\zeta$. Moreover, the mode $\theta$ from shift vector can be reduce to $\partial^2\theta/a^2\sim\epsilon\dot{\zeta}$ in this limit and can be also discard because it is subleading.
Combining all contributions finally the leading order of the interactions can be generally written as
\begin{equation}\label{interaction}
    \mathcal{L}_{\text{inv.dec}}= \frac{a^3\xi}{2}\left(\zeta+\frac{\lambda H }{M^2-\nabla^2/a^2}\dot{\zeta}\right)F_{\mu\nu}\tilde{F}^{\mu\nu}+\cdots,
\end{equation}
where the quantities of UV theory $\eta_{\bot}$ and $h$ have been absorbed into the dimensionaless function $\lambda=2h\eta_{\bot}$. Comparing with the single-field inflation we have a new vertex, which is the interaction between the $\dot{\zeta}$ and the EM fields. The strength of this coupling coefficient $\lambda$ is determined by the UV details of the EFT, for example, are the kinetic energy ratio $h$ and the turning rate $\eta_{\bot}$ in our case. Here we consider a general discussion and only assume the adiabatic condition $|\dot{\lambda}/\lambda|\ll |M|$ so that we can treat it as a free parameter in the following discussion. The magnitude of the parameter is related to the detials of the UV theory. But if the effects of heavy modes are important, we expect large $\eta_{\bot}$ and $h$, which means $\lambda$ could be not so small. We can naively assume that a smaller $c_s$ implies a larger $\lambda$. We will see interesting results in the parameter zone $c_s\ll1$ and $\lambda\gg 1$. In this paper we only consider $\lambda\gtrsim\mathcal{O}(1)$\footnote{One of the examples is the rapid-trun inflation \cite{Bjorkmo:2019fls} (sidetracked inflation \cite{Garcia-Saenz:2018ifx} and hyperbolic inflation \cite{Brown:2017osf} are specific cases of this kind of inflation), which give $\lambda\sim \eta_{\bot}^2\gg 1$. However, in this case the propagator could be quite different due to the imaginary speed of sound \cite{Garcia-Saenz:2018ifx,Fumagalli:2019noh,Garcia-Saenz:2018vqf}. We don't include discussion about this case in this paper.}.

The second interaction in (\ref{interaction}) can be rewritten through integral by parts as
\begin{align}\label{zetaAAhy}
    \mathcal{L}_{\text{inv.dec}}\supseteq-\frac{a^3}{2}\frac{\xi\lambda H}{M^2-\nabla^2/a^2}\zeta\partial_t\left(F_{\mu\nu}\tilde{F}^{\mu\nu}\right)+\partial_t\tilde{D},
\end{align}
where the boundary 
\begin{equation}
\tilde{D}=\frac{a^3\xi\lambda H }{2M^2-2\nabla^2/a^2}\zeta F_{\mu\nu}\tilde{F}^{\mu\nu}\propto a^3\boldsymbol{E}\cdot\boldsymbol{B}
\end{equation}
also do not contribute to correlation functions on super-horizon scales. We can see, if we write down the equation of motion of the curvature fluctuation, there are two kinds of source terms coming from the EM fields hence also have contributions to the observations. This effect arising due to inverse decay processes $\delta A+\delta A\rightarrow \zeta$ \cite{Barnaby:2010vf,Barnaby:2011vw}. However, the first kind of source is the same as \cite{Barnaby:2010vf,Barnaby:2011vw} while the second kind of source is the time derivative of Chern-Simon term. This is due to the coupling of EM fields with entropic fluctuation in the UV theory. In the low-energy effective theory, it reduces to this derivative coupling with curvature fluctuation.

\section{Correlation functions}
\label{sec:cf}

To concert the observations we calculate the power spectrum and non-Gaussianity of the curvature fluctuation during inflation. Due to the numerous particles production of the EM fields, we need to the consider the higher-order contributions to the curvature fluctuation from these EM fields. The correlation function can be computed by using the in-in formalism \cite{Maldacena:2002vr,Weinberg:2005vy}. Specifically, for an interaction Hamiltonian $H_I(t)$ in the interaction picture, the expectation of an operator $O(t)$, which is sandwiched by two ``in'' states, is given by
\begin{align}
    \langle\Omega| O(t) |\Omega\rangle=&\langle0| \left[\bar{T}\exp\left(i\int^{t}_{t_i}\ dt'H_I(t')\right)\right] O_I(t) \left[T\exp\left(-i\int^{t}_{t_i}\ dt''H_I(t'')\right)\right]|0\rangle\nonumber\\
    =&\sum_{n=0}^\infty (-i)^n\int_{t_i}^{t}dt_{1}\cdots\int_{t_i}^{t_{n-1}}dt_{n}\left\langle \left[\left[\left[O_I(t),H_I(t_1)\right],H_I(t_2)\right]\cdots,H_I(t_n)\right] \right\rangle,
\end{align}
where $|\Omega\rangle$ is the vacuum of the interacting theory and $|0\rangle$ is the one in the free theory, $T$($\bar{T}$) denotes the (anti-)time-ordered product. Introducing the $i\epsilon$ prescription at the initial time $t_i$ would switch off the excitation states and project the interaction vacuum into the free vacuum.
The interaction Hamiltonian here is at third order and obtained by cubic Lagrangian $H_I=-\int d^3x\ \mathcal{L}^{(3)}$. For the model in this paper, There are several contributions to the correlation function of the curvature fluctuation: the self-interactions of curvature fluctuation, the interaction of curvature fluctuation with the heavy field and the EM field.

\begin{figure}[tbp]
\centering
\includegraphics[scale=0.2]{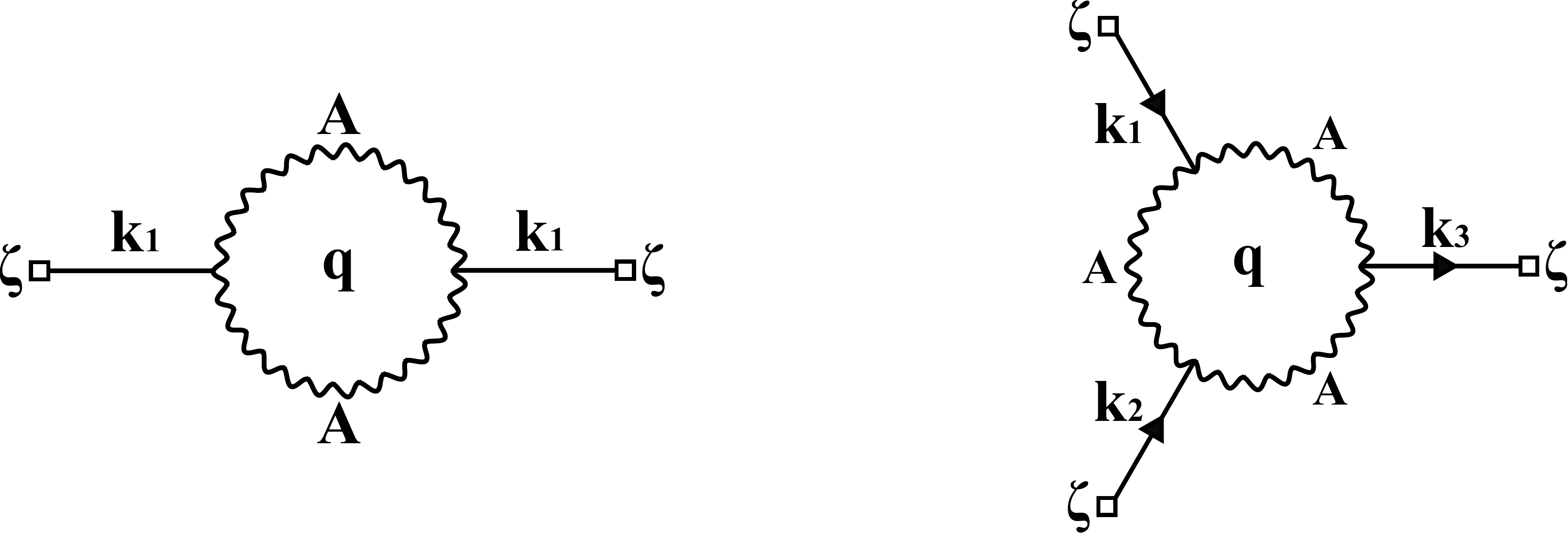}
\caption{\label{fig:vertices}The two-point and three-point correlation functions of inverse decay processes.}
\end{figure}

There are two kinds of interacting in $H_{\text{inv.dec}}=H_{\text{inv.dec}}^{(1)}+H_{\text{inv.dec}}^{(2)}$, coming from the interaction of EM fields with adiabatic and entropic fluctuations in the UV theory. For simply we still use the form (\ref{interaction}) to do the calculations. The two interactions of $H_{\text{inv.dec}}$ can be written in the form
\begin{align}
    H_{\text{inv.dec}}^{(1)}=&\sqrt{2\epsilon}M_{\text{pl}}\int\frac{d^3k}{(2\pi)^3}\ \zeta_{\boldsymbol{-k}}(t)J_{\boldsymbol{k}}(t),\label{H1}\\
    H_{\text{inv.dec}}^{(2)}=&\lambda H\sqrt{2\epsilon}M_{\text{pl}}\int\frac{d^3k}{(2\pi)^3}\ \frac{\dot{\zeta}_{\boldsymbol{-k}}(t)}{M^2+p^2}J_{\boldsymbol{k}}(t),\label{H2}
\end{align}
where
\begin{equation}
    J_{\boldsymbol{k}}(t)\equiv-\frac{a^3\xi}{2M_{\text{pl}}\sqrt{2\epsilon}}\int d^3x\ e^{-i\boldsymbol{k}\cdot\boldsymbol{x}}\left(F_{\mu\nu}\tilde{F}^{\mu\nu}-\langle F_{\mu\nu}\tilde{F}^{\mu\nu}\rangle\right)
\end{equation}
is the source from the EM fields to the curvature. We have subtracted the vacuum contributions and also the disconnected diagram such as $\langle J\rangle\langle J\rangle$ to the correlation functions. We want to compute the correlation functions of the curvature fluctuation $Q(t)=\zeta(t)\zeta(t)\cdots\zeta(t)$ at some specific time outside the horizon,
\begin{align}\label{cf}
    \langle \zeta_{\boldsymbol{k}_1}\zeta_{\boldsymbol{k}_2}\cdots \zeta_{\boldsymbol{k}_N}(t)\rangle_{\text{inv.dec}}
    =&\sum_{n=0}^{\infty}\left(-i\sqrt{2\epsilon}M_{\text{pl}}\right)^n\left[\prod_{k=1}^{n}\int_{t_i}^{t_k}dt_{k}\int\frac{d^3q_k}{(2\pi)^3}J_{\boldsymbol{q}_k}(t_k)\right]\nonumber\\
    &\times\langle\bigg[\bigg[\bigg[\zeta_{\boldsymbol{k}_1}\zeta_{\boldsymbol{k}_2}\cdots \zeta_{\boldsymbol{k}_N}(t),\zeta_{\boldsymbol{-q}_1}(t_1)+\frac{\lambda H\dot{\zeta}_{\boldsymbol{-q}_1}(t_1)}{M^2+p_1^2}\bigg],\nonumber\\
    &\ \ \ \ \zeta_{\boldsymbol{-q}_2}(t_2)+\frac{\lambda H\dot{\zeta}_{\boldsymbol{-q}_2}(t_2)}{M^2+p_2^2}\bigg]\cdots,\zeta_{\boldsymbol{-q}_n}(t_n)+\frac{\lambda H\dot{\zeta}_{\boldsymbol{-q}_n}(t_n)}{M^2+p_n^2}\bigg]\rangle,
\end{align}
where $\boldsymbol{p}_i\equiv \boldsymbol{q}_i/a(t_i)$ and we have used the relation (\ref{commu}) that all sources $J_{\boldsymbol{q}_k}(t_k)$ are commuting variables and we can take them out of the commutators\footnote{In \cite{Ballardini:2019rqh, Animali:2022lig}, similar results are displayed with gauge fields treated quantum mechanically with out commutation relation (\ref{commu}).}. Now the only thing we need to do is computing the commutator of the curvature fluctuations and correlation of the sources. Defining the normalized mode function
\begin{equation}
    v_{\boldsymbol{k}}\equiv-a\sqrt{2\epsilon}M_{\text{pl}}\zeta_{\boldsymbol{k}},
\end{equation}
then for $t_1>t_2$ we have 
\begin{align}
    [\zeta_{\boldsymbol{k}_1}(t_1),\zeta_{\boldsymbol{k}_2}(t_2)]
    =\frac{-i(2\pi)^3}{2\epsilon M_{\text{pl}}^2}\frac{G_{k_1}(t_1,t_2)}{a_1a_2}\delta\left(\boldsymbol{k}_1+\boldsymbol{k}_{2}\right),\\
    [\zeta_{\boldsymbol{k}_1}(t_1),\dot{\zeta}_{\boldsymbol{k}_2}(t_2)]
    =\frac{-i(2\pi)^3}{2\epsilon M_{\text{pl}}^2}\frac{F_{k_1}(t_1,t_2)}{a_1a_2}\delta\left(\boldsymbol{k}_1+\boldsymbol{k}_{2}\right),
\end{align}
where $a_n\equiv a(t_n)$ and functions $G_k(t,t')$ and $F_k(t,t')$ read
\begin{align}
    G_k(t,t')\equiv &i\theta(t-t')\left[v_{k}(t)v_{k}^*(t')-v_{k}^*(t)v_{k}(t')\right],\label{rGreen}\\
    F_k(t,t')\equiv &i\theta(t-t')a(t')\left[v_{k}(t)\left(\frac{v_{k}^*(t')}{a(t')}\right)^.-v_{k}^*(t)\left(\frac{v_{k}(t')}{a(t')}\right)^.\right],\label{rFreen}
\end{align}
which are given by the mode functions of the curvature fluctuation in the EFT (\ref{EFTUV}). $G_k(t,t')$ is the retarded Green function. After inserting this commutator into (\ref{cf}) and using the mode functions of the EM fields (\ref{Asol}) we can compute the correlation functions of the curvature fluctuation. For example, For example, in this paper we calculate the two-point and three-point correlation of inverse decay processes shown as Figure \ref{fig:vertices}. From now on we denote $J_i\equiv J_{\boldsymbol{k}_i}(\tau_i)$ for simply.

\subsection{Power spectrum}
\label{sec:PS}
We first compute the two-point correlation function $N=2$ for inverse decay processes. We drop the subscript ``onv.dec'' in this section for simple. The $n=0$ term is the Gaussian contribution that nearly scale-invariant of the fluctuation. It's known that the number of Gaussian fields is odd in $n=1$ term hence its contribution is vanishing. The leading nonvanishing term is given by $n=2$,
\begin{align}\label{zetaAA2}
    \langle \zeta_{\boldsymbol{k}_1}\zeta_{\boldsymbol{k}_2}\rangle=&
    \left(\frac{-1}{\sqrt{2\epsilon}M_{\text{pl}}}\right)^2\int_{\tau_i}^{\tau}d\tau_1d\tau_2\ \Bigg[\frac{G_{k_1}(\tau,\tau_1)}{a(\tau)}\frac{G_{k_2}(\tau,\tau_2)}{a(\tau)}\nonumber\\
    &+2\frac{\lambda H}{M^2+p_2^2}\frac{G_{k_1}(\tau,\tau_1)}{a(\tau)}\frac{F_{k_2}(\tau,\tau_2)}{a(\tau)}\nonumber\\
    &+\frac{\lambda H}{M^2+p_1^2}\frac{\lambda H}{M^2+p_2^2}\frac{F_{k_1}(\tau,\tau_1)}{a(\tau)}\frac{F_{k_2}(\tau,t_2)}{a(\tau)}\Bigg]\langle J_1J_2\rangle.
\end{align}
We have used the fact that $J_1J_2\sim\delta(\boldsymbol{k}_1+\boldsymbol{k}_2)$. To compute the correlation function $\langle J_1J_2\rangle$, we should insert the quantized EM fields (\ref{qA}) into the $F_{\mu\nu}\tilde{F}^{\mu\nu}$. It's obvious that the only non-vanishing contributions are the $\langle\hat{a}_{-}\hat{a}_{-}^{\dagger}\hat{a}_{-}\hat{a}_{-}^{\dagger}\rangle$ and the $\langle\hat{a}_{-}\hat{a}_{-}\hat{a}_{-}^{\dagger}\hat{a}_{-}^{\dagger}\rangle$ terms. The former term is the disconnected diagram $\langle J\rangle\langle J\rangle$ and we have discarded it. Only the later term is left. 
we need to insert the mode function, which is given by (\ref{Asol}). In the case that all interesting modes of curvature are in the interval $(8\xi )^{-1}\lesssim-k\tau\lesssim 2\xi $. The correlation $\langle JJ\rangle$ is evaluated to (\ref{JJf}). 
On the other hand, we also need to compute the function of curvature fluctuation (\ref{rGreen}) and (\ref{rFreen}) and we should discuss them in different regimes.

\subsubsection{Linear dispersion regime}
In this regime the linear part of the dispersion relation is dominated. The effect of heavy modes is reflected in the reduced speed of sound. We can use the low-energy EFT (\ref{linearEFT}), where the equation of motion of curvature fluctuation in the de Sitter background is given
\begin{equation}
    \ddot{\zeta}_k+3H\dot{\zeta}_k+\frac{c_s^2k^2}{a^2}\zeta_k=0,
\end{equation}
The solution with Bunch-David vacuum initial conditions reads
\begin{equation}\label{zetasol}
    \zeta_k=\frac{H}{2\sqrt{\epsilon c_s}M_{\text{pl}}k^{3/2}}(1+ikc_s\tau)e^{-ikc_s\tau},
\end{equation}
where $\tau$ is the conformal time. The propagation of curvature fluctuation is the same as single-field case except the smaller speed of sound so we can directly adopt most of the results from \cite{Barnaby:2011vw}. We are always interested in the correlation of $\zeta$ on super-horizon scale where $-k\tau\rightarrow0$. 
The contribution from vacuum modes is well-known as the following power spectrum
\begin{align}\label{lps}
     \mathcal{P}_{\text{(L)}}=&\frac{H^2}{8\pi^2\epsilon c_s M_{\text{pl}}^2},
\end{align}
where we have assumed the de Sitter limit and drop the tilts of the power spectrum.

Now we consider the inverse decay contributions. Inserting (\ref{JJf}) and (\ref{zetasol}) into (\ref{zetaAA2}) we obtain the first contribution at late time
\begin{align}
     \langle \zeta_{\boldsymbol{k}_1}\zeta_{\boldsymbol{k}_2}\rangle_{GG}
     =&\frac{2\pi^2}{k_1^3}\mathcal{P}_{\text{(L)}}^2f_2^{GG}(\xi)e^{4\pi\xi }(2\pi)^3\delta(\boldsymbol{k}_1+\boldsymbol{k}_2),
\end{align}
The dimensionaless function $f_2^{GG}$ is evaluated by (\ref{f21}). One can rescale the variable $x_c\equiv c_s x$ in the integrals and find that this is equivalent to rescaling the parameter $\xi_c\equiv\xi c_s^{-1}$ in $f_2^{GG}$. Hence the approximation (\ref{Asol}) used in the whole integration region in (\ref{f21}) is also valid \cite{Barnaby:2011vw}. This is physically, due to a sound horizion-crossing scale of curvature modes. Rescaling the variable implies the rescale of the spatial coordinate hence restoring the (fake) Lorentz invariance of the action (\ref{zeta2}). The causality ensures the location of the peak $|\boldsymbol{q}_*|\sim \mathcal{O}(1)$ in the integration (\ref{f21}) is the same order as horizon-crossing scale $x_c\sim \mathcal{O}(1)$. Hence we need a larger $\xi_c>\xi$ to produce the same contributions as $c_s=1$ case on a sound horizon length. Then the integral $f_2^{GG}$ can be approximately calculated by
\begin{align}\label{f2A}
    f^{GG}_2(\xi)\simeq 
   \frac{7.5\times 10^{-5}}{\xi^{6}}c_s^6,\ \ \ \ \ \ \ \ \xi\gg c_s.
\end{align}

Similarly, we can calculate the second and the third term in (\ref{zetaAA2}). In the linear dispersion regime, the momentum $p\ll M$ hence can be ignored. Then these contributions can be calculated to
\begin{align}
     \langle \zeta_{\boldsymbol{k}_1}\zeta_{\boldsymbol{k}_2}\rangle_{GF}
     &=2\lambda \left(\frac{H}{M}\right)^2\frac{2\pi^2}{k_1^3}\mathcal{P}_{\text{(L)}}^2 f_2^{GF}e^{4\pi\xi }(2\pi)^3\delta(\boldsymbol{k}_1+\boldsymbol{k}_2),\\
    \langle \zeta_{\boldsymbol{k}_1}\zeta_{\boldsymbol{k}_2}\rangle_{FF}
     &=\lambda^2\left(\frac{H}{M}\right)^4\frac{2\pi^2}{k_1^3}\mathcal{P}_{\text{(L)}}^2 f_2^{FF}e^{4\pi\xi }(2\pi)^3\delta(\boldsymbol{k}_1+\boldsymbol{k}_2),
\end{align}
where $f_2^{GF}$ and $f_2^{FF}$ are given by (\ref{f22}). We note that for large $z$, the function $f_2^{GF}(\xi)\simeq -3f_2^{GG}(\xi)$ and $f_2^{FF}(\xi)\simeq 9f_2^{GG}(\xi)$. Then we can use
\begin{equation}
\begin{aligned}
     f_2^{GF}(\xi)&\simeq 
   -\frac{2.3\times 10^{-4}}{\xi^{6}}c_s^6,\\
     f_2^{FF}(\xi)&\simeq 
   \frac{6.8\times 10^{-4}}{\xi^{6}}c_s^6,
\end{aligned}\ \ \ \ \ \ \ \ \xi\gg c_s
\end{equation}
to approximately calculate. Because we are interested in the small sound speed case $c_s\ll 1$, the approximation is valid for the interesting regime $\xi\gtrsim 1$. In conclusion, the total contributions of the inverse decay process is given by
\begin{align}\label{twopf}
     \langle \zeta_{\boldsymbol{k}_1}\zeta_{\boldsymbol{k}_2}\rangle
     =&\frac{2\pi^2}{k_1^3}\mathcal{P}_{\text{(L)}}^2e^{4\pi\xi }(2\pi)^3\delta(\boldsymbol{k}_1+\boldsymbol{k}_2)\\
     &\times\left[f_2^{GG}(\xi)+2\lambda\left(\frac{H}{M}\right)^2f_2^{GF}(\xi)+\lambda^2\left(\frac{H}{M}\right)^4 f_2^{FF}(\xi)\right].
\end{align}
We see that $f^{GF}_2$, 
$f^{GG}_2$ and $f^{FF}_2$ scale in the same way. Therefore the magnitude of these new vertexes compared to the first one read
\begin{equation}
    \frac{\langle \zeta\zeta\rangle_{GF}}{\langle \zeta\zeta\rangle_{GG}}\simeq -6.1\lambda\left(\frac{H}{M}\right)^2, \ \ \ \ \ \ \ 
    \frac{\langle \zeta\zeta\rangle_{FF}}{\langle \zeta\zeta\rangle_{GG}}\simeq 9.1\lambda^2\left(\frac{H}{M}\right)^4.
\end{equation}
The contributions from the second and the third term are suppressed by the mass of heavy field. However, they are also related to the quantity $\lambda$, which could be a large number in the low-energy EFT. Hence for fixing $\Lambda_{\text{UV}}$ when $c_s$ is small enough, which implies low enough $M\lesssim H$, the second and the third term could be significant. We will discuss it in the following subsection.

We note that the second contribution $\langle\zeta\zeta\rangle_{GF}$ is negative if $\lambda>0$. In the parameter zone $0.27\lesssim\lambda(H/M)^2\lesssim0.4$ the inverse decay process can reduce the total amplitude of the power spectrum. For $\lambda(H/\Lambda_{\text{UV}})^2\ll 1$, this is the region corresponding to small speed of sound $c_s\ll 1$.% We show the parameter space $c_s-\xi$ as Figure \ref{fig:negative} for $\Lambda_{\text{UV}}=100H$ and $\xi=3$ as example. The blue region denotes the negative contribution of $\langle\zeta\zeta\rangle_{\text{inv.dec}}$. The dashed line is given by $2\xi H=M$. The l.h.s of this line is the region where the non-linear dispersion regime should be considered. 
The negative contribution is danger because the inverse decay is exponentially sensitive to $\xi$ and may result in a veary lager value. The parameter zone of negative contribution is summarized in Table \ref{tab:negative}. We will discuss their contributions to the power spectrum and consider the constrains from the observation in section \ref{sec:ip}.

\begin{table}[t]
\centering
\renewcommand\arraystretch{1.7}
 \setlength{\tabcolsep}{2.1mm}{
\begin{tabular}{c|c|c}
\hline
        & Linear dispersion & Non-linear dispersion \\ \hline
$\langle\zeta\zeta\rangle_{\text{inv.dec}}<0$ & $0.27< \lambda\left(\frac{H}{M}\right)^2< 0.4$              & $\times$                   \\ \hline
$\langle\zeta\zeta\zeta\rangle_{\text{inv.dec}}<0$ & $\lambda\left(\frac{H}{M}\right)^2> 0.43$               & $\lambda\xi^2>1.3$  \\ \hline                
\end{tabular}
}
\caption{Negative correlation function.}
\label{tab:negative}
\end{table}

\subsubsection{Non-linear dispersion regime}
If the $M$ is low enough so that the particle production is in the non-linear dispersion regime, where we need to consider the EFT Lagrangian (\ref{EFTUV}), the propagation of modes should be modified and also the $p^2$ in interaction (\ref{H2}) becomes important. The equation of motion is given by
\begin{align}\label{nllimit}
    \ddot{\zeta}_k+3H\dot{\zeta}_k+\frac{2H(1-c_s^2)M^2p^2}{(M^2+p^2)(M^2+c_s^2p^2)}\dot{\zeta}_k+\omega^2\zeta_k=0,
\end{align}
where $\omega$ is given by (\ref{dispersion}). We have an other scale-dependent term and a $\omega^2$ term in the equation. The coefficients of these terms are generally time-dependent because of $p=k/a$ hence the equation cannot be solved analytically. Here we consider a limit case,  where $c_s\ll 1$ and the modes are deeply in the non-linear dispersion regime,
\begin{equation}\label{nldl}
    M^2\ll p^2\ll M^2c_s^{-2}.
\end{equation}
The particle production of EM fields is also during the non-linear dispersion regime. In this limit, the equation of motion is simplified to
\begin{equation}
    \ddot{\zeta}_k+5H\dot{\zeta}_k+\frac{k^4}{a^4\Lambda_{\text{UV}}^2}\zeta_k=0.
\end{equation}
To solve this equation, note that we should impose a modified initial conditions on sub-horzion scales because of the modified commutation relation from (\ref{EFTUV}) (see details in appendix \ref{ad1}). The solution is \cite{Baumann:2011su,Gwyn:2012mw}
\begin{equation}\label{nlzetasol}
    \zeta_k=i\frac{H^2(-\tau)^{5/2}}{\sqrt{2\epsilon}M_{\text{pl}}}\sqrt{\frac{\pi}{8}}\frac{k}{\Lambda_{\text{UV}}}H^{(1)}_{5/4}(u), \ \ \ \ \ \ u=\frac{H}{2\Lambda_{\text{UV}}}k^2\tau^2,
\end{equation}
where $H^{(1)}_{\nu}$ is the Hankel function of the first kind. 

\begin{figure}[tbp]
\centering
\includegraphics[scale=0.7]{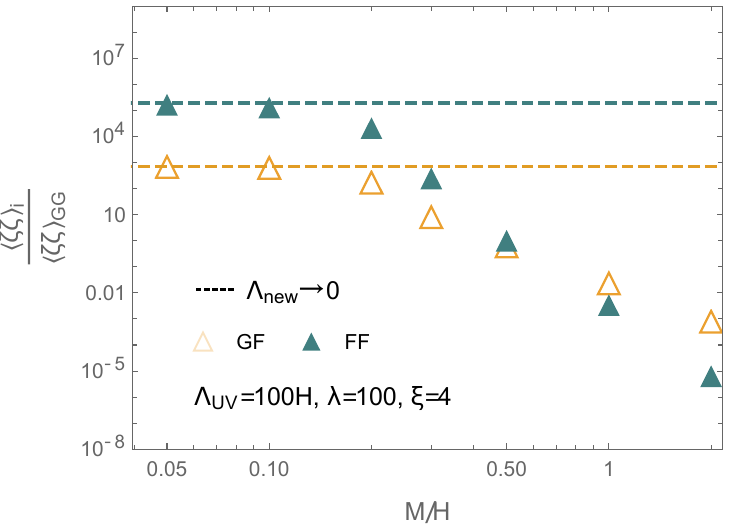}
\caption{\label{fig:I3I2I1}Comparing $\langle\zeta\zeta\rangle_{GF}$ and $\langle\zeta\zeta\rangle_{FF}$ with $\langle\zeta\zeta\rangle_{GG}$ by fixing the UV scale and changing the mass of heavy field. Their contributions become significant when $M\lesssim H$ and saturates at the dashed lines, which are evaluated by (\ref{dashline}).}
\end{figure}

In the correlation function, contribution of the second interaction has a factor $(M^2+p^2)^{-1}$ in each time integral. The momentum become more and more important when entering the non-linear dispersion regime $p^2\sim M^2$. We can roughly calculate the two-point correlation by assuming the $sudden$ transition between the non-linear and linear dispersion regime. The transition time $\tau_{\text{new}}$ is given by $p(\tau_{\text{new}})=p_{\text{new}}=M$. Before $\tau_{\text{new}}$, where the modes are deeply in the non-linear dispersion regime, the mode solutions are given by (\ref{nlzetasol}). After $\tau_{\text{new}}$, the modes enter the linear dispersion regime and the propagation is described by
\begin{align}
    \zeta_k=\frac{HA_{+}}{\sqrt{2\epsilon c_s}M_{\text{pl}}k^{3/2}}(1+ikc_s\tau)e^{-ikc_s\tau}+\frac{HA_{-}}{\sqrt{2\epsilon c_s}M_{\text{pl}}k^{3/2}}(1-ikc_s\tau)e^{ikc_s\tau}.
\end{align}
To determine the coefficients of positive and negative frequencies $A_{\pm}$, we impose the boundary conditions of these two regimes
\begin{equation}
    \zeta(\tau_{\text{new}}^{-})=\zeta(\tau_{\text{new}}^{+}),\ \ \ \ \ \ \ \ p_{\zeta}(\tau_{\text{new}}^{-})=p_{\zeta}(\tau_{\text{new}}^{+}),
\end{equation}
where $p_{\zeta}\equiv\partial\mathcal{L}/\partial\zeta'$ is the conjugate momentum of $\zeta$. Note that in the non-linear dispersion regime, we use (\ref{nldl}) hence the momentum boundary condition implies $\zeta'(\tau_{\text{new}}^{-})=\zeta'(\tau_{\text{new}}^{+})$. Then the coefficients are given by
\begin{align}
    A_{\pm}=\mp\sqrt{\frac{\pi}{32}}\left(\frac{\Lambda_{\text{new}}}{H}\right)^{1/2}\left[H^{(1)}_{5/4}(u_{\text{new}})-\left(\frac{H}{\Lambda_{\text{new}}}\pm i\right)H^{(1)}_{1/4}(u_{\text{new}})\right]e^{\mp i\frac{\Lambda_{\text{new}}}{H}},
\end{align}
where $u_{\text{new}}=\Lambda_{\text{new}}/(2H)$. We use these mode functions to numerically calculate three contributions in (\ref{zetaAA2}). Figure \ref{fig:I3I2I1} shows the magnitude of $GF$ and $FF$ contributions compared to the $GG$ one by fixing UV cut-off $\Lambda_{\text{UV}}=100H$ and coupling with EM fileds $\xi=4$ and $\lambda=100$, and then changing the new physics scale $M$. We see when the mass of heavy modes is greater than Hubble scale, the second contribution is small and can be discarded. Once the new scale $M/H\lesssim1$, the $GF$ and $FF$ contributions quickly become more and more important, and even becomes greater than the $GG$ one. This ratio saturates for small enough $M$ or $\Lambda_{\text{new}}$. Let's  discuss this limit case in the following. 

The extremely low $\Lambda_{\text{new}}$ case, where the horizon-crossing take places during the non-linear dispersion regime: $\Lambda_{\text{new}}\ll H\ll\Lambda_{\text{UV}}$, from equation of motion the horizon-crossing takes places at $p\sim H\Lambda_{\text{UV}}$, consistent with the domination of no-linear $p^4$ in dispersion relation (\ref{nld}) when $\omega=H$. In this low-energy limit, The contribution from vacuum modes is given by the power spectrum
\begin{align}\label{nlps}
    \mathcal{P}_{(\text{NL})}=\frac{\Gamma^2(5/4)H^2}{\pi^3\epsilon M_{\text{pl}}^2}\sqrt{\frac{\Lambda_{\text{UV}}}{H}}.
\end{align}

Then the first contribution of the interactions to the correlation is given by
\begin{align}\label{zeta21n}
     \langle \zeta_{\boldsymbol{k}_1}\zeta_{\boldsymbol{k}_2}\rangle_{GG}
     =&\frac{2\pi^2}{k_1^3}\mathcal{P}_{(\text{NL})}^2g_2^{GG}(\xi)e^{4\pi\xi }(2\pi)^3\delta(\boldsymbol{k}_1+\boldsymbol{k}_2),
\end{align}
where the function $g_2^{GG}$ is given by (\ref{g2GG}). Similar to the linear dispersion case, we can redefine the parameter $\xi_{\text{UV}}\equiv\xi\sqrt{2\Lambda_{\text{UV}}/H}$ in the correlation function. The horizon-crossing take places at a smaller length $\sqrt{\Lambda_{\text{UV}}H}\gg H$ and we need a larger effective parameter $\xi_{\text{UV}}\gg\zeta$ to produce sufficient gauge fields in $g_2^{GG}$ due to the causality. 
Then the dimensionaless function $g_2^{GG}$ is evaluates to
\begin{equation}\label{g2}
    g_2^{GG}(\xi)\simeq \frac{2.1\times10^{-3}}{\xi^{10}}\left(\frac{H}{\Lambda_{\text{UV}}}\right)^5,\ \ \ \ \ \ \ \xi\gg \sqrt{\frac{H}{2\Lambda_{\text{UV}}}}.
\end{equation}
Because $H/\Lambda_{\text{UV}}$ is a small number, for the regime of EM field we are interested in $\xi\gtrsim 1$ the condition $\xi\gg \sqrt{H/\Lambda_{\text{UV}}}$ is always satisfied. Hence the approximation (\ref{g2}) works well in this regime, as Figure \ref{fig:g2tot} shows. 

For the second and the third contributions, in the non-linear dispersion limit (\ref{nldl}) the mass $M$ can be ignored compared to the momentum $p$. Then Using the derivative of Hankel function (\ref{dH1}), the function $F_k(\tau,\tau')$ can be calculated and we finally have
\begin{align}
     \langle\zeta_{\boldsymbol{k}_1}\zeta_{\boldsymbol{k}_2}\rangle_{GF}
     =&2\lambda\left(\frac{H}{\Lambda_{\text{UV}}}\right)\frac{2\pi^2}{k_1^3}\mathcal{P}_{(\text{NL})}^2 g^{GF}_2(\xi)e^{4\pi\xi }(2\pi)^3\delta(\boldsymbol{k}_1+\boldsymbol{k}_2),\label{zeta22n}\\
     \langle \zeta_{\boldsymbol{k}_1}\zeta_{\boldsymbol{k}_2}\rangle_{FF}
     =&\lambda^2\left(\frac{H}{\Lambda_{\text{UV}}}\right)^2\frac{2\pi^2}{k_1^3}\mathcal{P}_{(\text{NL})}^2 g^{FF}_2(\xi)e^{4\pi\xi }(2\pi)^3\delta(\boldsymbol{k}_1+\boldsymbol{k}_2),\label{zeta12n}
\end{align}
where the functions $g_2^{GF}$ and $g_2^{FF}$ are evaluated by (\ref{g2GF}) and (\ref{g2FF}), and approximately results in
\begin{equation}
\begin{aligned}    
    g^{GF}_2(\xi)&\simeq -\frac{4.6\times 10^{-4}}{\xi^8}\left(\frac{H}{\Lambda_{\text{UV}}}\right)^4,\\
    g^{FF}_2(\xi)&\simeq \frac{1.6\times 10^{-4}}{\xi^6}\left(\frac{H}{\Lambda_{\text{UV}}}\right)^3,
\end{aligned}\ \ \ \ \ \ \ \xi\gg \sqrt{\frac{H}{2\Lambda_{\text{UV}}}}.
\end{equation}
We also show the agreement of this approximation and numerical calculation in Figure \ref{fig:g2tot} in the interesting regime. We see in this non-linear dispersion limit, the contribution of the inverse decay process is dependent on the UV scale $\Lambda_{\text{UV}}$ rather than $M$. In conclusion, the two-point correlation of inverse decay process is 
\begin{align}\label{twopf2}
     \langle \zeta_{\boldsymbol{k}_1}\zeta_{\boldsymbol{k}_2}\rangle
     =&\frac{2\pi^2}{k_1^3}\mathcal{P}_{(\text{NL})}^2e^{4\pi\xi }(2\pi)^3\delta(\boldsymbol{k}_1+\boldsymbol{k}_2)\nonumber\\
     &\times\left[g_2^{GG}(\xi)+2\lambda\left(\frac{H}{\Lambda_{\text{UV}}}\right)g_2^{GF}(\xi)+\lambda^2\left(\frac{H}{\Lambda_{\text{UV}}}\right)^2g_2^{FF}(\xi)\right].
\end{align}
The total amount of these contributions are suppressed by the UV scale $\Lambda_{\text{UV}}$. However, we still have a factor $\lambda$, which should be a large number.

We note that the $g^{GF}_2$ and $g^{FF}_2$ have factors $(H/\Lambda_{\text{UV}})^4$ and $(H/\Lambda_{\text{UV}})^3$ respectively. And from (\ref{zeta22n}) and (\ref{zeta12n}) we find the correlation function of these interaction include a factor $(H/\Lambda_{\text{UV}})^5$, which is the same as the $g_2^{GG}$ and hence as the first contribution (\ref{zeta21n}). Also, the 
$g^{GF}_2$ and $g^{FF}_2$ fall more slowly than $g_2^{\text{A}}$ with increasing $\xi$. Their saturated ratio for fixed $\Lambda_{\text{UV}}$ are evaluated by
\begin{equation}\label{dashline}
    \frac{\langle \zeta\zeta\rangle_{GF}}{\langle \zeta\zeta\rangle_{GG}}\bigg{|}_{\Lambda_{\text{new}}\to 0}\simeq -0.44\lambda\xi^2, \ \ \ \ \ \ \ 
    \frac{\langle \zeta\zeta\rangle_{FF}}{\langle \zeta\zeta\rangle_{GG}}\bigg{|}_{\Lambda_{\text{new}}\to 0}\simeq 0.076\lambda^2\xi^4, 
\end{equation}
which are independent on any cut-off scale of the EFT, but depends on the magnitude of inverse decay $\xi$. We show these two ratios as dashed lines in Figure \ref{fig:I3I2I1}. These two contributions is equal to the first one at $|\xi^2\lambda|\simeq 2.3$ and $3.6$ respectively. For not so small $\lambda\gtrsim \mathcal{O}(1)$, the contributions from the second inverse decay can become greater than the first one in the regime $\xi\gtrsim 1$. Similarly to the linear dispersion regime, the term $\langle\zeta\zeta\rangle_{GF}$ also contributes negatively if $\lambda>0$. However we found the total inverse decay contributions is positive for any $\lambda$ and $\xi$. We will discuss their total contributions to the power spectrum in section \ref{sec:ip}.

\subsection{Bispectrum}
\label{sec:bispectrum}
We next calculate the three-point correlation function $N=3$ of the curvature fluctuation. To discuss the shape of the triangle formed by momentum, we denote the amplitudes of three momentum as
\begin{align}
    |\boldsymbol{k}_1|=k,\ \ \ \ \ \ \ \ |\boldsymbol{k}_2|=x_2k,\ \ \ \ \ \ \ \ |\boldsymbol{k}_3|=x_3k.
\end{align}
For inverse decay processes, the largest contributions to the three-point correlation function is from the equilateral configuration $x_2=x_3=1$, due to the peak contribution from the EM field are all around $|\boldsymbol{k}_i|\sim\mathcal{O}{(|\boldsymbol{q}|)}$ \cite{Barnaby:2010vf,Barnaby:2011vw}, where $\boldsymbol{q}$ is the internal momentum of the EM field. Therefore we here only consider the equilateral contribution of the correlation function. The leading non-vanishing contribution is 
\begin{align}
    \langle \zeta_{\boldsymbol{k}_1}\zeta_{\boldsymbol{k}_2}\zeta_{\boldsymbol{k}_3}\rangle^{\text{equil}}=&
    \left(\frac{-1}{\sqrt{2\epsilon}M_{\text{pl}}c_s^2}\right)^3\int_{\tau_i}^{\tau}d\tau_1d\tau_2d\tau_3\ \Bigg[\frac{G_{k_1}(\tau,\tau_1)}{a(\tau)}\frac{G_{k_2}(\tau,\tau_2)}{a(\tau)}\frac{G_{k_3}(\tau,\tau_3)}{a(\tau)}\nonumber\\
    &+3\frac{\lambda H}{M^2+p_3^2}\frac{G_{k_1}(\tau,\tau_1)}{a(\tau)}\frac{G_{k_2}(\tau,\tau_2)}{a(\tau)}\frac{F_{k_3}(\tau,\tau_3)}{a(\tau)}\nonumber\\
    &+3\frac{\lambda H}{M^2+p_2^2}\frac{\lambda H}{M^2+p_3^2}\frac{G_{k_1}(\tau,\tau_1)}{a(\tau)}\frac{F_{k_2}(\tau,\tau_2)}{a(\tau)}\frac{F_{k_3}(\tau,\tau_3)}{a(\tau)}\nonumber\\
    &+\frac{\lambda H}{M^2+p_1^2}\frac{\lambda H}{M^2+p_2^2}\frac{\lambda H}{M^2+p_3^2}\frac{F_{k_1}(\tau,\tau_1)}{a(\tau)}\frac{F_{k_2}(\tau,\tau_2)}{a(\tau)}\frac{F_{k_3}(\tau,\tau_3)}{a(\tau)}\Bigg]\nonumber\\
    &\times\langle J_1J_2J_3\rangle.
\end{align}
The correlation $\langle J_1J_2J_3\rangle$ is evaluated to (\ref{JJJf}) by using the approximation of gauge potential solutions in de Sitter limit. We also discuss the linear and non-linear dispersion regime respectively.

\subsubsection{Linear dispersion regime}
For the linear dispersion regime, the three-point correlation of equilateral configuration of the inverse decay process is given by
\begin{align}\label{bis1}
    \langle \zeta_{\boldsymbol{k}_1}\zeta_{\boldsymbol{k}_2}\zeta_{\boldsymbol{k}_3}\rangle^{\text{equil}}=&\frac{3(2\pi)^{7}}{10}\mathcal{P}^3\cdot 3e^{6\pi\xi }\frac{\delta (\boldsymbol{k}_1+\boldsymbol{k}_2+\boldsymbol{k}_3)}{k^6}\Bigg[f_3^{GGG}(\xi;1,1)\nonumber\\
    &+3\lambda\left(\frac{H}{M}\right)^2f_3^{GGF}(\xi;1,1)+3\lambda^2\left(\frac{H}{M}\right)^4f_3^{GFF}(\xi;1,1)\nonumber\\
    &+\lambda^3\left(\frac{H}{M}\right)^6f_3^{FFF}(\xi;1,1)\Bigg].
\end{align}
The dimensionaless function of contributions from the first interaction is (\ref{f3A}). Also, after a rescaling of the variable $x_c=c_sx$ in the integral. It's equivalent to rescaling of the parameter $\xi_c=\xi c_s^{-1}$ in $f_3^{\text{A}}$. Therefore we can directly adopt the results from \cite{Barnaby:2011vw}. We are interested in the equilateral configuration $x_2=x_3=1$. By using the integral $\mathcal{I}_{\text{E}}$ (\ref{f22}) and recalling $\mathcal{I}_{\text{E}}\simeq -3\mathcal{I}_{\text{A}}$ for large $\xi c_s^{-1}$, the inverse decay in the equilateral configuration are
\begin{equation}
\begin{aligned}
    f^{GGG}_3(\xi;1,1)&\simeq 
   \frac{2.8\times 10^{-7}}{\xi^{9}}c_s^9,\\
    f^{GGF}_3(\xi;1,1)&\simeq 
   -\frac{8.3\times 10^{-7}}{\xi^{9}}c_s^9,\\
   f^{GFF}_3(\xi;1,1)&\simeq 
   \frac{2.5\times 10^{-6}}{\xi^{9}}c_s^9,\\
    f^{FFF}_3(\xi;1,1)&\simeq -
   \frac{7.5\times 10^{-6}}{\xi^{9}}c_s^9,
\end{aligned}
   \ \ \ \ \ \ \ \ \xi\gg c_s.
\end{equation}
As the same as the two-point correlation, all these term scale in the same way. Then the contributions of the second inverse decay are associated with one $\lambda(H/M)^2$ compared to the first one if we have one more ``$F$'' integral in the correlation function.  We found that in the linear dispersion regime $M\gg 2\xi H$, the total contributions of inverse decay to the three-point correlation is negative if $\lambda(H/M)^2>0.43$. Hence for small enough speed of sound and at the same time large $\lambda$, the inverse decay reduces the magnitude of three-point correlation.

\subsubsection{Non-linear dispersion regime}
For the non-linear dispersion regime, we also consider the non-linear dispersion limit (\ref{nldl}) for analytically computable and use the modes function (\ref{nlzetasol}). The contribution of equilateral configuration from inverse decay (\ref{interaction}) is given by
\begin{align}\label{bis2}
    \langle \zeta_{\boldsymbol{k}_1}\zeta_{\boldsymbol{k}_2}\zeta_{\boldsymbol{k}_3}\rangle^{\text{equil}}=&\frac{3(2\pi)^{7}}{10}\mathcal{P}^3\cdot 3e^{6\pi\xi }\frac{\delta (\boldsymbol{k}_1+\boldsymbol{k}_2+\boldsymbol{k}_3)}{k^6}\Bigg[g_3^{GGG}(\xi;1,1)\nonumber\\
    &+3\lambda\left(\frac{H}{\Lambda_{\text{UV}}}\right)g_3^{GGF}(\xi;1,1)+3\lambda^2\left(\frac{H}{\Lambda_{\text{UV}}}\right)^2g_3^{GFF}(\xi;1,1)\nonumber\\
    &+\lambda^3\left(\frac{H}{\Lambda_{\text{UV}}}\right)^{3}g_3^{FFF}(\xi;1,1)\Bigg].
\end{align}
The contribution of the first interaction is given by (\ref{g3A}). After using the approximation (\ref{JA}) and considering the equilateral configuration, we obtain
\begin{equation}
\begin{aligned}
    g_3^{GGG}(\xi;1,1)&\simeq \frac{1.5\times 10^{-4}}{\xi^{15}}\left(\frac{H}{\Lambda_{\text{UV}}}\right)^{15/2},\\
    g_3^{GGF}(\xi;1,1)&\simeq -\frac{5.8\times 10^{-5}}{\xi^{13}}\left(\frac{H}{\Lambda_{\text{UV}}}\right)^{13/2},\\
    g_3^{GFF}(\xi;1,1)&\simeq \frac{1.8\times 10^{-5}}{\xi^{11}}\left(\frac{H}{\Lambda_{\text{UV}}}\right)^{11/2},\\
    g_3^{FFF}(\xi;1,1)&\simeq -\frac{6.5\times 10^{-6}}{\xi^{9}}\left(\frac{H}{\Lambda_{\text{UV}}}\right)^{9/2}.
\end{aligned}
     \ \ \ \ \ \ \ \xi\gg \sqrt{\frac{H}{2\Lambda_{\text{UV}}}}.
\end{equation}
Similarly to the two-point correlation, all of these terms contributing to the three-point correlation are $\sim(H/\Lambda_{\text{UV}})^{15/2}$. Hence the hierarchy of these contributions is controlled by $\lambda\xi^2$. Moreover, the contributions to the $f_{\text{NL}}^{\text{equil}}$ is negative if $\lambda\xi^2>1.3$. Therefore in almost the entire parameter zone we are interested in of the non-linear dispersion regime, the inverse decay reduces the magnitude of non-Gaussianity.

\section{Example: sidetracked inflation}
We here show the sidetracked inflation of a two-scalar system $(\phi,\psi)$ \cite{Garcia-Saenz:2018ifx} as example. We consider the case whose metric of the field space is a hyperbolic plane
\begin{equation}
    G_{IJ}d\phi^Id\phi^J=\left(1+\frac{2\psi^2}{M_{\text{fs}}^2}\right)d\phi^2+\frac{2\sqrt{2}\psi}{M_{\text{fs}}}d\phi d\psi+d\psi^2,
\end{equation}
which has a constant scalar curvature $R_{\text{fs}}=-4/M_{\text{fs}}^2$. Here $M$ is the curvature scale of the field space and is sub-Planckian $M_{\text{fs}}\lesssim M_{\text{pl}}$. On the other hand, the potential of this system has form
\begin{equation}
    V(\phi,\psi)=\Lambda^4\mathcal{V}(\phi)+\frac{1}{2}m_h^2\psi^2,
\end{equation}
where the dimensionless potential in pseudo-scalar inflation is given by $\mathcal{V}=1-\cos(\phi/f)$ and the mass of pseudo-scalar field is $m_a^2\simeq\Lambda^2/f^2$. The $\psi$ field is heavy here $m_h\gtrsim H$ so that the geometrical destabilization of single-field inflation occurs \cite{Renaux-Petel:2015mga}. The expansion of universe is driven by the pseudo-scalar hence we have $3H^2M_{\text{pl}}^2\simeq \Lambda^4\mathcal{V}$. 

\begin{figure}[tbp]
\centering
\includegraphics[scale=0.67]{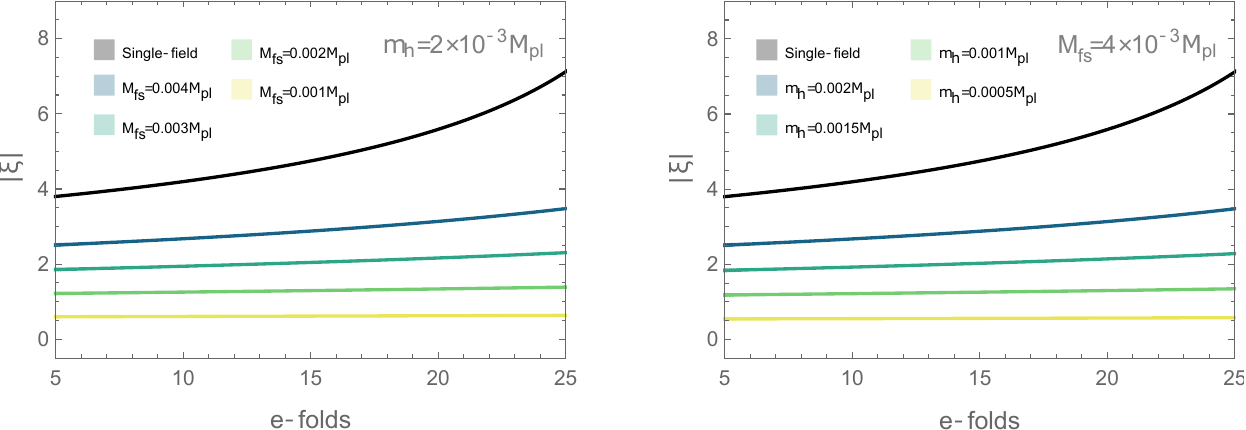}
\caption{\label{fig:zeta}$m_a=10^{-5}M_{\text{pl}}$, $\alpha=40$, $f=100$. Evolution of parameter $|\xi|$ for different $M_{\text{fs}}$(Left) and different $m_h$(Right).}
\end{figure}

The equations of motion of two scalar fields are
\begin{align}
    &\ddot{\phi}+3H\dot{\phi}+\frac{4\psi}{M_{\text{fs}}^2}\dot{\psi}\dot{\phi}+\frac{\sqrt{2}}{M_{\text{fs}}}\dot{\psi}^2+\frac{\sqrt{2}\psi}{M_{\text{fs}}}\left[2\frac{\dot{\phi}^2}{M_{\text{fs}}^2}\psi-V_{\psi}\right]+V_{\phi}=0,\\
    \ddot{\psi}+3H\dot{\psi}&-\frac{\psi}{M_{\text{fs}}^2}\dot{\psi}^2-\frac{4\sqrt{2}\psi^2}{M_{\text{fs}}^3}\dot{\phi}\dot{\psi}+\left(1+\frac{2\psi^2}{M_{\text{fs}}^2}\right)\left[-2\frac{\dot{\phi}^2}{M_{\text{fs}}^2}\psi+V_{\psi}\right]-\sqrt{2}\frac{\psi}{M_{\text{fs}}}V_{\phi}=0.
\end{align}
The evolution of $\psi$ is much slower than $\phi$ hence we can neglect the time derivative of $\psi$. Then equation of motion of $\psi$ is reduced to
\begin{align}\label{phieomST}
    2\frac{\dot{\phi}^2}{M_{\text{fs}}^2}\simeq m_h^2-\sqrt{2}\frac{V_{\phi}}{M_{\text{fs}}}\left(1+\frac{2\psi^2}{M_{\text{fs}}^2}\right)^{-1}.
\end{align}
Inserting this equation into the equation of motion of $\phi$, we obtatin
\begin{align}\label{chieom}
    3H\dot{\phi}\simeq -V_{\phi}\left(1+\frac{2\psi^2}{M_{\text{fs}}^2}\right)^{-1}.
\end{align}
Combining two equations (\ref{phieomST}) and (\ref{chieom}) and in the heavy limit $m_h\gg H$ one can derive
\begin{align}\label{phieomlo}
    \frac{\dot{\phi}}{H}\simeq -\frac{M_{\text{fs}}m_h}{\sqrt{2}H}.
\end{align}
This implies that the second term in the r.h.s of (\ref{phieomST}) is sub-leading in this limit. So now it's clear that the parameter 
\begin{equation}
    \xi\equiv \frac{\alpha\dot{\phi}}{2fH}\simeq -\frac{\alpha}{2\sqrt{2}}\frac{M_{\text{fs}}m_h}{fH}.
\end{equation}
The scale $M_{\text{fs}}$ is the curvature scale of field space hence if $M_{\text{fs}}$ is large enough, the field space becomes flat and the $\phi$ decouples with $\psi$. In this limit, the impact of heavy field is not important to the inflation anymore. On the other hand, if the heavy field is heavy enough, it's difficult to excite it at low-energy scales. The critical value is given by $M_{\text{fs}}m_h=V_{\phi}/(3H)$. We show the slowly rolling $\xi$ with different $M_{\text{fs}}$ and $m_h$ in Figure. \ref{fig:zeta}.

\begin{figure}[tbp]
\centering
\includegraphics[scale=0.67]{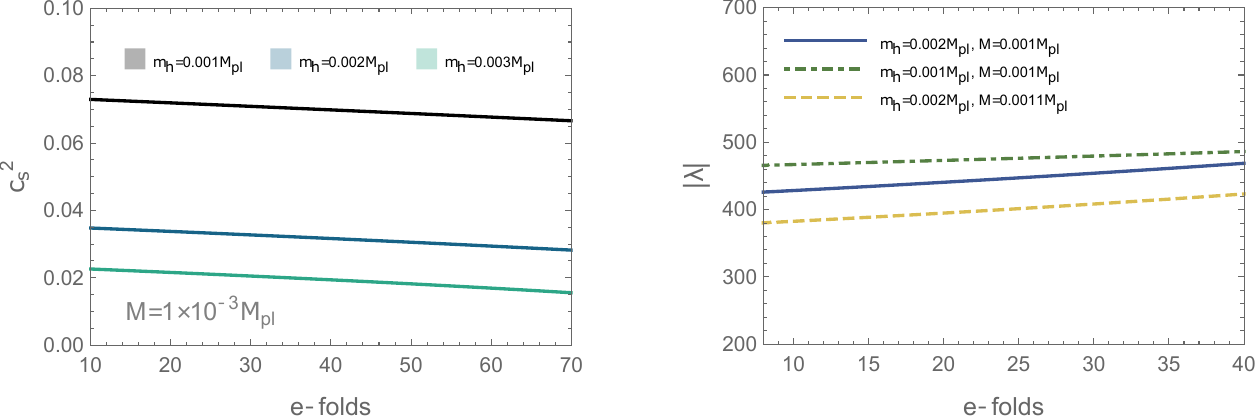}
\caption{\label{fig:Mmh}$m_a=10^{-5}M_{\text{pl}}$. Left: Speed of sound for different $m_h$. Right: Coupling $|\lambda|$ for different $m_h$ and $M_{\text{fs}}$.}
\end{figure}

We next can use (\ref{phieomST}) and (\ref{chieom}) to derive the turning-rate
\begin{equation}
    \eta_{\bot}\equiv-\frac{V_N}{H\dot{\sigma}}\simeq-\frac{2\psi}{M_{\text{fs}}^2}\frac{\dot{\phi}}{H}
\end{equation}
and the mass of entropic fluctuation $M$
\begin{equation}
    M^2\simeq 2\sqrt{2}\frac{2\psi^2V_{\phi}}{M_{\text{fs}}^3}\left(1+\frac{2\psi^2}{M_{\text{fs}}^2}\right)^{-1}.
\end{equation}
Notice that we use the (\ref{phieomST}) rather than the leading-order (\ref{phieomlo}) when calculate the $M$. Then we can use this leading-order to approximately reduce the turning-rate and entropic mass to
\begin{equation}
    \eta_{\bot}\simeq \frac{\sqrt{2}\psi}{M_{\text{fs}}}\frac{m_h}{H},\ \ \ \ \ \ \ \ \ M^2\simeq 12\frac{\psi^2}{M_{\text{fs}}^2}m_hH,
\end{equation}
The small speed of sound $c_s\ll 1$ can be estimated as
\begin{equation}
    c_s^2\simeq \frac{3H}{2m_h},
\end{equation}
which is only dependent on the mass of heavy field $m_h$. We show the slowly rolling $c_s$ with different $m_h$ in Figure \ref{fig:Mmh}.

To obtain the parameter $\lambda$, we also need to derive $h$. To do this, we transform the field space metric to $d\phi^2+e^{-2\sqrt{2}\phi/M_{\text{fs}}}d\chi^2$ through $e^{-\sqrt{2}\phi/M_{\text{fs}}}\chi= \psi$. Then if we ignore the evolution of $\psi$, we obtain $h\simeq -\sqrt{2}\psi/M_{\text{fs}}$ and
\begin{equation}
    \lambda\equiv 2h\eta_{\bot}\simeq -\frac{2\psi^2}{M_{\text{fs}}^2}\frac{m_h}{H}\simeq -2\sqrt{8\pi\epsilon_V}\frac{M_{\text{pl}}}{M_{\text{fs}}}+\frac{m_h}{H},
\end{equation}
where in the last equality we have used (\ref{chieom}), (\ref{phieomlo}) and defined $\epsilon_V\equiv(M_{\text{pl}}^2/16\pi)(V_{\phi}/V)^2=M_{\text{pl}}^2/(4\pi\phi^2)$. We can learn that $|\lambda|\gg 1$ is a large parameter due to $\lambda\sim M^2/H^2$, as we expected. We can see, heavier $m_h$ reduces the value $|\lambda|$, while heavier $M_{\text{fs}}$ enhances it. We show these results in Figure \ref{fig:Mmh}.

After deriving $\xi$, $M$, $c_s$ and $\lambda$, we can discuss how heavy fields influence the primordial power spectrum and bispectrum in this model. We have found formula of the contributions of non-local vertices, which were discussed in linear and non-linear limit. 
\subsection{Linear dispersion}
\begin{figure}[tbp]
\centering
\includegraphics[scale=0.75]{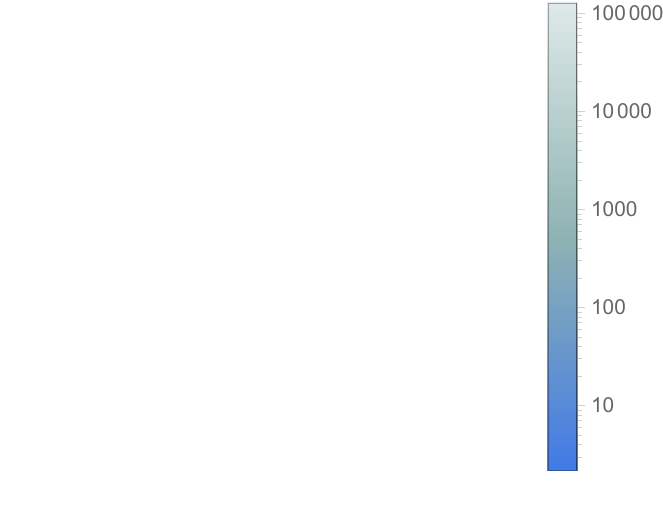}
\caption{\label{fig:MathcalG}Function $\mathcal{G}(\alpha M_{\text{fs}}/f,m_h/H)$. We only interested in the $|\xi|>1$ case, which is shown above the red curve.}
\end{figure}

For linear dispersion, we have known in (4.21) that all contributions to the power spectrum scale as $(\xi/c_s)^{-6}$. Then we can obtain the ratio between the second or third therms and the first term
\begin{align}
    \left|\frac{\langle \zeta\zeta\rangle_{\text{non-local}}}{\langle \zeta\zeta\rangle_{GG}}\right|\sim \left|\lambda\left(\frac{H}{M}\right)\right|^{2n}\simeq \left(\frac{1}{6}\right)^{n},
\end{align}
where $n=1$ for $GF$ term and $n=2$ for $FF$ term. We found that this ratio is not suppressed by heavy entropic mass $M$, it is a not very small number. Then the non-local contributions are also important in the linear-dispersion regime. The same goes for equilateral bispectrum. The non-local terms have the same order as the first term in the bispectrum. Combining these, the total contributions of inverse decay to two-point and three-point correlations are estimated as
\begin{align}
    &\langle \zeta\zeta\rangle_{\text{inv.dec}}\simeq \frac{2\pi^2}{k_1^3}\mathcal{P}^2(2\pi)^3\times0.29\ \mathcal{G}^2\left(\frac{\alpha M_{\text{fs}}}{f}, \frac{m_h}{H}\right),\\
    &\langle \zeta\zeta\zeta\rangle_{\text{inv.dec}}\simeq\frac{9(2\pi)^{7}}{10k^6}\mathcal{P}^3\times 0.067\ \mathcal{G}^3\left(\frac{\alpha M_{\text{fs}}}{f}, \frac{m_h}{H}\right),
\end{align}
where we define a function $\mathcal{G}$ as
\begin{equation}
    \mathcal{G}(x,y)\equiv\exp{\left[\frac{\pi}{\sqrt{2}}x\cdot y \right]} x^{-3}\cdot y^{-9/2}
\end{equation}
We show the density plot of $\mathcal{G}(\alpha M_{\text{fs}}/f,m_h/H)$ in Figure (\ref{fig:MathcalG}).

\subsection{Non-linear dispersion}
For non-linear dispersion, we have learned from (4.37) that the ratio of non-local and the first contributions to correlations are characterized by $\lambda\xi^2$. This is a very large number hence we can only approximately calculate the $FF$ term and $FFF$ terms in the correlations. The UV cut-off in this model is given by
\begin{equation}
    \Lambda_{\text{UV}}\equiv Mc_s^{-1}\simeq 2\sqrt{2}\frac{\psi}{M_{\text{fs}}}m_h.
\end{equation}
Then the contributions of inverse decay to two-point and three-point correlations are estimated as
\begin{align}
    &\langle \zeta\zeta\rangle_{\text{inv.dec}}\simeq \frac{2\pi^2}{k_1^3}\mathcal{P}^2(2\pi)^3\times0.0026\ \mathcal{G}^2\left(\frac{\alpha M_{\text{fs}}}{f}, \frac{m_h}{H}\right)\left(\frac{M_{\text{fs}}}{\sqrt{2}\psi}\right),\\
    &\langle \zeta\zeta\zeta\rangle_{\text{inv.dec}}\simeq-\frac{9(2\pi)^{7}}{10k^6}\mathcal{P}^3\times 0.0004\ \mathcal{G}^3\left(\frac{\alpha M_{\text{fs}}}{f}, \frac{m_h}{H}\right)\left(\frac{M_{\text{fs}}}{\sqrt{2}\psi}\right)^{3/2}.
\end{align}
These results are very similar to the linear-dispersion case, except a factor $M_{\text{fs}}/(\sqrt{2}\psi)$. The heavy field $\psi$ is excited due to the hyperbolic geometry of field space hence we always expect that $\psi\sim M_{\text{fs}}$. Therefore it usually gives us a order unit factor.

\section{Implications on phenomenology}
\label{sec:ip}
\subsection{Scalar amplitude}

\begin{figure}[tbp]
\centering
\includegraphics[scale=0.75]{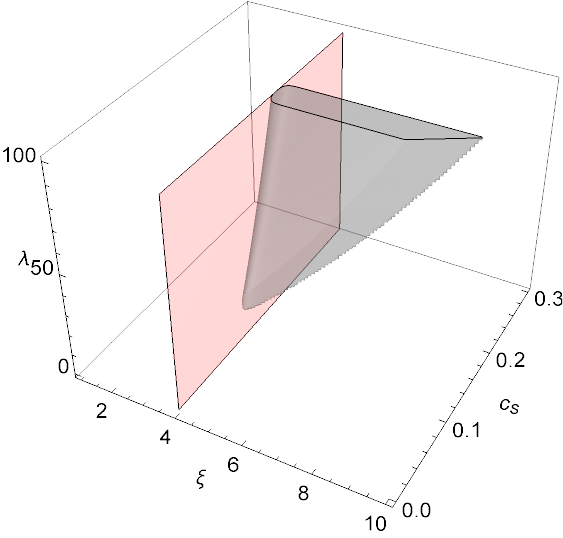}
\caption{\label{fig:realps}$\Lambda_{\text{UV}}=100H$. The parameter zone of imaginary $\mathcal{P}_{\text{Planck}}$ in the regime $2\xi H<M$ (gray). The red plane is the $\xi=4$ plane. }
\end{figure}

In this last section we have calculate the two-point correlation of two contributions of the inverse decay processes. We address the total contributions to the scalar amplitude by fixing $\Lambda_{\text{UV}}$. Combining these two contributions and also the free single-field contribution, the primordial power spectrum in linear and non-linear dispersion limit at late time reads
%\begin{align}
%    P_{\zeta(\text{L})}=&\mathcal{P}_{\text{(L)}}\left(\frac{k}{k_0}\right)^{n_s-1}\left[1+\left(f_2^{\text{A}}(\xi)+\frac{\lambda^2H^4}{M^4}f^{\text{E}}_2(\xi)\right)\mathcal{P}_{\text{(L)}}e^{4\pi\xi }\right],\\
%    P_{\zeta\text{(NL)}}=&\mathcal{P}_{\text{(NL)}}\left(\frac{k}{k_0}\right)^{n_s-1}\left[1+\left(g_2^{\text{A}}(\xi)+\frac{\lambda^2H^2}{\Lambda_{\text{UV}}^2} g^{\text{E}}_2(\xi)\right)\mathcal{P}_{\text{(NL)}}e^{4\pi\xi }\right].
%\end{align}
\begin{equation}
    P_{\zeta}=\mathcal{P}\left[1+\mathcal{P}f_2^{\text{tot}}(\xi)e^{4\pi\xi}\right],
\end{equation}
where
\begin{align}\label{PStot}
     f_{2}^{\text{tot}}(\xi)=\left\{
     \begin{aligned}
     &f_2^{GG}+\frac{2\lambda H^2}{\Lambda_{\text{UV}}^2c_s^2}f^{GF}_2+\frac{\lambda^2H^4}{\Lambda_{\text{UV}}^4c_s^4}f^{FF}_2,\ \ &M\gg 2\xi H\\
     &g_2^{GG}+\frac{2\lambda H}{\Lambda_{\text{UV}}} g^{GF}_2+\frac{\lambda^2H^2}{\Lambda_{\text{UV}}^2} g^{FF}_2,\ \ &M\ll H
     \end{aligned}
     \right.
\end{align}
The amplitude of the total scalar fluctuations is given by Planck2018 result $P_{\zeta}=2.1\times 10^{-9}$ \cite{Planck:2018vyg}. Note that for these two limits the vacuum power spectrum $\mathcal{P}$ are given by (\ref{lps}) and (\ref{nlps}) respectively. Then the amplitude of the vacuum parts can be represent as
\begin{equation}\label{PPlanck}
    \mathcal{P}_{\text{Planck}}(\xi)=\frac{e^{-4\pi\xi}}{2f_2^{\text{tot}}(\xi)}\left[-1+\sqrt{1+8.4\times 10^{-9}f_2^{\text{tot}}(\xi)e^{4\pi\xi}}\right].
\end{equation}
Firstly, the vacuum power spectrum should be real so $1+8.4\times 10^{-9}f_2^{\text{tot}}(\xi)e^{4\pi\xi}>0$. We have known in the non-linear dispersion limit $f_2^{\text{tot}}$ is always positive. Hence we should look at the linear dispersion regime $M>2\xi H$. We show this constrain in Figure \ref{fig:realps} for $\Lambda_{\text{UV}}=100H$. We see that there exists a large parameter space can avoid unphysical $\mathcal{P}_{\text{Planck}}$.

Moreover, we should ensure that for the interesting parameter space, there is no any backreaction to the inflation. The conditions are provided in appendix \ref{appendix:be}. It's useful to discuss it at $\xi_{\text{eq}}$, which is defined as the value that the vacuum and inverse decay have the same contribution to the power spectrum, i.e. $\mathcal{P}_{\text{Planck}}(\xi_{\text{eq}})=1.05\times10^{-9}$. 
We discuss how $\xi_{\text{eq}}$ relies on the effects of the heavy modes by fixing the scale $\Lambda_{\text{UV}}$. By using the power spectrum, we have
\begin{align}\label{xieq}
   \xi_{\text{eq}}^{3}e^{-2\pi\xi_{\text{eq}}}\simeq 
   \left\{\begin{aligned}
   &0.009c_s^3\sqrt{\left|1-6.1\lambda\left(\frac{H}{\Lambda_{\text{UV}}c_s}\right)^2+9.1\lambda^2\left(\frac{H}{\Lambda_{\text{UV}}c_s}\right)^4\right|},\ \ \ \ \ &M\gg 2\xi_{\text{eq}}H\\
   &0.05\left(\frac{H}{\Lambda_{\text{UV}}}\right)^{5/2}\sqrt{\frac{1}{\xi_{\text{eq}}^{4}}-0.4\frac{\lambda}{\xi_{\text{eq}}^2}+0.08\lambda^2},\ \ \ \ \ &M\ll H
   \end{aligned}
   \right.
\end{align}
In the non-linear dispersion limit, we are interested in $\xi_{\text{eq}}\gg 1$ and $\lambda\gg 1$, in which the third term in the root is dominated. Hence we can ignored the first and the second terms for simple. 

\begin{figure}[tbp]
\centering
\includegraphics[scale=1]{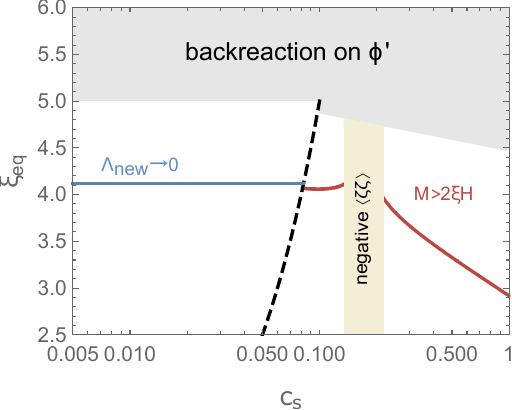}
\caption{\label{fig:xieq}$\Lambda_{\text{UV}}=100H$, $\lambda=100$ and $\mathcal{A}=10$. In the non-linear dispersion limit, the $\xi_{\text{eq}}$ only depends on the UV scale hence is a constant if we fix $\Lambda_{\text{UV}}$, as the blue curve shows. In the linear regime it depends on $c_s$, as the red curve shows. For $c_s=1$ and $\lambda^2H^4/\Lambda_{\text{UV}}^4\ll 1$ the second inverse decay is negligible so we go back to the result $\xi_{\text{eq}}\simeq 2.9$ in \cite{Barnaby:2011vw}. The gray region is the strong backreaction of the EM field to the axion. The yellow region has negative power spectrum.}
\end{figure}

On the other hand, we need to avoid the backraction effects from electromagnetic fields on $\phi$, which is descirbed by (\ref{phibe}). We should note that due to the dynamics of heavy fields, we have $\epsilon=\epsilon_{\phi}+K^2\epsilon_{\chi}$. We naturally expect that the two terms on the r.h.s has the same order hence $\epsilon\gtrsim \epsilon_{\phi}$. For example, for three case in l.h.s of Fiugre. \ref{fig:Mmh}, we have $\mathcal{A}\sim 1+2\psi^2/M_{\text{fs}}^2\simeq 12,\ 6,\ 4$ respectively. By using (\ref{lps}), (\ref{nlps}) the boundaries constrains of backreaction effect are given by
\begin{equation}\label{xiback}
    1.5\times 10^{11}\mathcal{A}\xi_{\text{eq,back}}^{3}e^{-2\pi\xi_{\text{eq,back}}}=\left\{
     \begin{aligned}
     &c_s,\ \ \ \ \ &M\gg 2\xi_{\text{eq,back}} H\\
     &\frac{\pi}{8\Gamma^2[5/4]}\left(\frac{H}{\Lambda_{\text{UV}}}\right)^{1/2},\ \ \ \ \ &M\ll H
     \end{aligned}
     \right.
\end{equation}
where $\mathcal{A}\sim \epsilon/\epsilon_{\phi}\gtrsim\mathcal{O}(1)$. The boundaries are different in linear and non-linear dispersion regime due to the different power spectrum parametrization of these two regimes. Because $\xi^3e^{-2\pi\xi}$ is a monotonically decreasing function, there is no strong backreaction, i.e., $\xi_{\text{eq}}<\xi_{\text{eq,back}}$ if the r.h.s of (\ref{xieq}) is greater than the r.h.s of (\ref{xiback}). It's easily satisfied in a wide range of parameter space due to the factor $10^{11}$ on the l.h.s of (\ref{xiback}), which comes from the amplitude of the scalar fluctuation. We show the $\xi_{\text{eq}}$-$c_s$ in Figure \ref{fig:xieq} for an example $\Lambda_{\text{UV}}=100H$, $\lambda=100$ and $\mathcal{A}=10$. The black dashed curve divides the linear and non-linear dispersion regime, which is linear on the r.h.s while non-linear on the l.h.s. We show (\ref{xieq}) as red and blue curves, which are in the safe zone of parameter space.

\subsection{Non-Gaussianity}
The non-Gaussianity can be characterized by a dimensionaless parameter $f_{\text{NL}}$, defined by expanding perturbations around Gaussianity in position space
\begin{align}
    \zeta(\boldsymbol{x})=\zeta_g(\boldsymbol{x})+\frac{3}{5}f_{\text{NL}}\left[\zeta_g^2(\boldsymbol{x})-\langle\zeta_g^2(\boldsymbol{x})\rangle\right],
\end{align}
where $\zeta_g$ is the Gaussian field. In the momentum space,
\begin{align}
    \zeta_{\boldsymbol{k}}=\zeta_{\boldsymbol{k}}^g+\frac{3}{5}f_{\text{NL}}\int \frac{d^3q}{(2\pi)^3}\ \zeta_{\boldsymbol{q}}^g\zeta_{\boldsymbol{q}-\boldsymbol{k}}^g.
\end{align}
Then the three-point correlation function can be represented through two-point correlation functions of Gaussian fields as
\begin{align}
    \langle\zeta_{\boldsymbol{k}_1}\zeta_{\boldsymbol{k}_2}\zeta_{\boldsymbol{k}_3}\rangle=\frac{3(2\pi)^7}{10}f_{\text{NL}}P_{\zeta}^2\frac{1+x_2^3+x_3^3}{x_2^3x_3^3}\frac{\delta (\boldsymbol{k}_1+\boldsymbol{k}_2+\boldsymbol{k}_3)}{k^6}.
\end{align}
The parameter $f_{\text{NL}}$ characterizes the magnitude of the non-Gaussianity. There are several contributions to this parameter in the low-energy EFT: the self-interaction of curvature fluctuation $\zeta$ and the inverse decay processes (\ref{interaction}). It's known that the reduced speed of sound implies the large interactions, giving raise to enhancement of the non-Gaussinity through $\dot{\zeta}(\partial\zeta)^2$ and $\dot{\zeta}^3$. 
\begin{figure}[tbp]
\centering
\includegraphics[scale=0.75]{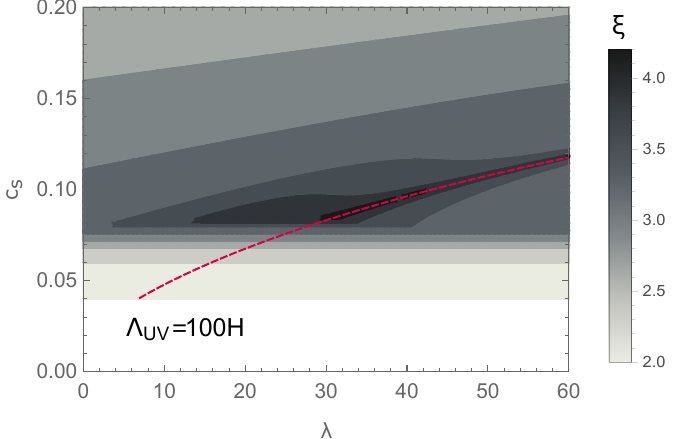}
\caption{\label{fig:fNL100}Linear dispersion: $\Lambda_{\text{UV}}=100H$. The parameter space of $-100\lesssim (f_{\text{NL}}^{\text{equil}})_{\text{inv.dec}}\lesssim 100$ for various $\xi$. The red dashed curve divides the postive and negative value, i.e., above(below) the red dashed curve $(f_{\text{NL}}^{\text{equil}})_{\text{inv.dec}}>0(<0)$. We found this constrains the parameter $\xi_{\text{linear dispersion}}\lesssim 4.2$.}
\end{figure}

Now we consider the contributions from interactions (\ref{interaction}), due to the particles production of EM fields. By Using the results we derived in section \ref{sec:bispectrum}, the largest contribution to the non-Gaussianity is the equilateral configuration $|\boldsymbol{k}_1|=|\boldsymbol{k}_2|=|\boldsymbol{k}_3|$. Then we obtain
\begin{equation}
    \left(f_{\text{NL}}^{\text{equil}}\right)_{\text{inv.dec}}=\frac{f_3^{\text{tot}}(\xi;1,1)\mathcal{P}_{\text{Planck}}^3e^{6\pi\xi}}{P_{\zeta}^2}
\end{equation}
where
\begin{align}\label{Bstot}
     f_{3}^{\text{tot}}(\xi)=\left\{
     \begin{aligned}
     &f_3^{GGG}+\frac{3\lambda H^2}{\Lambda_{\text{UV}}^2c_s^2}f^{GGF}_3+\frac{3\lambda^2H^4}{\Lambda_{\text{UV}}^4c_s^4}f^{GFF}_3+\frac{\lambda^3H^6}{\Lambda_{\text{UV}}^6c_s^6}f^{FFF}_3,\ \ &M\gg 2\xi H\\
     &g_3^{GGG}+\frac{3\lambda H}{\Lambda_{\text{UV}}} g^{GGF}_3+\frac{3\lambda^2H^2}{\Lambda_{\text{UV}}^2} g^{GFF}_3+\frac{\lambda^3H^3}{\Lambda_{\text{UV}}^3} g^{FFF}_3,\ \ &M\ll H
     \end{aligned}
     \right.
\end{align}
Also, The terms with odd ``$F$'' have negative contributions if $\lambda>0$. Hence the sign of $(f_{\text{NL}}^{\text{equil}})_{\text{inv.dec}}$ is determined by $f_3^{\text{tot}}$ and $\mathcal{P}$. We have discussed the sign of $f_3^{\text{tot}}$ last section and has been summarized in Table \ref{tab:negative}. In linear dispersion regime, the constrain is only on the $\lambda-c_s$ surface hence we find there is no any overlap between negative $\langle\zeta\zeta\rangle$ and $\langle\zeta\zeta\zeta\rangle$. Hence the sign of non-Gaussianity is only determined by $f_3^{\text{tot}}$. It's worth mentioning that in the non-linear dispersion limit, $(f_{\text{NL}}^{\text{equil}})_{\text{inv.dec}}$ is negative in almost the entire parameter space we are interested in if $\lambda>0$ and vice versa. On the other hand, in the linear dispersion regime, the sign of this parameter is dependent on the parameters $\lambda(H/M)^2$.
\begin{figure}[tbp]
\centering
\includegraphics[scale=0.75]{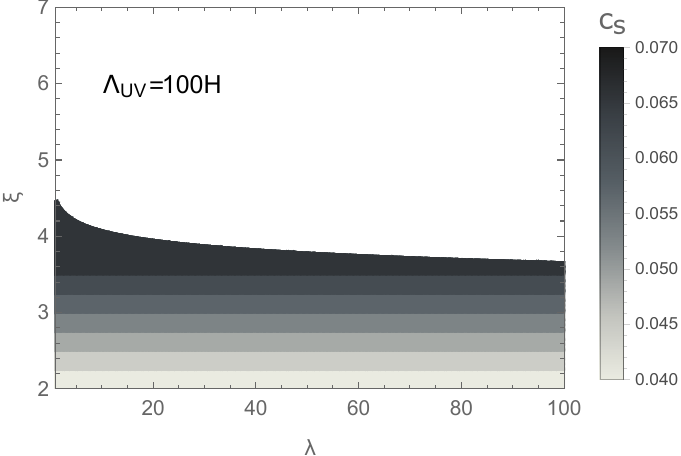}
\caption{\label{fig:fNL1002}Non-inear dispersion: $\Lambda_{\text{UV}}=100H$. The parameter space of $-100\lesssim (f_{\text{NL}}^{\text{equil}})_{\text{inv.dec}}\lesssim 0$ for various $c_s$. We found this constrains the parameter $\xi_{\text{non-linear dispersion}}\lesssim 4$ for a wide range of $\lambda$. The speed of sound provide the lower bound of $\xi$.}
\end{figure}

If we consider the large $\zeta$, the exponential term in (\ref{PPlanck}) is dominated. Also, in the non-linear dispersion limit, the $\xi^{-6}$ in $g_2^{\text{tot}}$ and $\xi^{-9}$ in $g_3^{\text{tot}}$ become dominated in this limit. Then by using the scalar amplitude from obseravtion $P_{\zeta}=2.1\times10^{-9}$ and the vacuum power spectrum (\ref{PPlanck}), the non-Gaussianity in two different regimes are reduced to
\begin{align}
    \left(f_{\text{NL}}^{\text{equil}}\right)_{\text{inv.dec}}^{\xi\to\infty}\simeq 21822\times\left\{
     \begin{aligned}
     &\frac{0.28-2.5X+7.5X^2-7.5X^3}{\left(0.75-4.6X+6.8X^2\right)^{3/2}},\ \ &M\gg 2\xi H\\
     &-\text{sign}\left(\lambda\right)3.2,\ \ &M\ll H
     \end{aligned}
     \right.
\end{align}
where $X\equiv \lambda(H/M)^2$. They saturate to a constant value at large $\zeta$. In the linear dispersion regime this value is determined by $\lambda(H/M)^2$, while in non-linear dispersion limit $(f_{\text{NL}}^{\text{equil}})_{\text{inv.dec}}^{\xi\to\infty}\simeq-\text{sign}(\lambda)69830$ does not depend on any parameters other than the sign of $\lambda$. In the linear dispersion regime, it approaches to $(f_{\text{NL}}^{\text{equil}})_{\text{inv.dec}}^{\xi\to\infty}\simeq-9255$ for larger $\lambda(H/M)^2$ and $9407$ for small $\lambda(H/M)^2$. These values, however, have exceeded the 68$\%$ CL limit of the Planck data $-21\lesssim f_{\text{NL}}^{\text{equil}}\lesssim 73$ \cite{Planck:2019kim}. In the linear dispersion regime, the non-Gaussianity from vacuum is enhanced by $(f_{\text{NL}}^{\text{equil}})_{\text{vac}}\sim 1/c_s^2$, while in non-linear dispersion regime $(f_{\text{NL}}^{\text{equil}})_{\text{vac}}\sim\Lambda_{\text{UV}}/H$ \cite{Gwyn:2012mw}. The UV scale is expected to be above symmetry breaking scale so $\Lambda_{\text{UV}}\sim 100H$. Then we can see in Figure \ref{fig:xieq}, the linear dispersion regime is about $c_s\gtrsim 0.1$. Hence we expect the non-Gaussianity from vacuum part of the curvature fluctuation is $(f_{\text{NL}}^{\text{equil}})_{\text{vac}}\sim 100$. Hence we also constrain the parameter space of the contribution from inverse decay process as $-100\lesssim (f_{\text{NL}}^{\text{equil}})_{\text{inv.dec}}\lesssim 100$ in Figure \ref{fig:fNL100} for linear dispersion regime. We found this limit rules out $\xi_{\text{linear dispersion}}>4.2$. For non-linear dispersion limit, although the upper bound of $\xi$ depends on $\lambda$, we found it is not really sensitive, as Figure \ref{fig:fNL1002} show. Then this rules out $\xi_{\text{non-linear dispersion}}\gtrsim 4$ for a wide range of $\lambda$. We find in the Figure \ref{fig:realps}, the parameter zone $\xi<4$ avoids unphysical areas of power spectrum.

\section{Conclusions and outlook}

In this study, we analyze the impact of heavy fields on pseudo-scalar inflation. Even for scalar perturbation modes with momentum $p$ below the energy scale $Mc_s^{-1}$, the non-linear dispersion relation plays a crucial role, provided that the momentum is greater than the mass of the heavy field. When the condition $Mc_s^{-1} > 2\xi p > M$ is satisfied near Hubble crossing, specifically when $p \approx H$, the effects from the non-linear dispersion relation contribute to the correlation functions of curvature perturbations. Within the framework of the effective field theory (EFT) of inflation, the interaction between the gauge field and curvature perturbation comprises two terms at the leading order. In addition to the conventional term $\zeta F\tilde F$, we identify a nesting term associated with the effects arising from the heavy fields. This nesting term leads to higher order derivative operators in the EFT \eqref{interaction}. We use the in-in formalism to calculate the curvature perturbation power spectrum of the model by considering the inverse decay processes at the first order.  We found that contributions of the second inverse decay to the power spectrum becomes significant than the first one when the mass of heavy fields is under Hubble scale $M\lesssim H$. Besides, by comparing with Planck data on power spectrum, we found a wide range of parameter space for avoiding large back-reaction effects and strong coupling regime in which the perturbation theory is not valid. We also calculate the equilateral non-Gaussianity in the two dispersion regimes. We found that for a wide range of $\lambda$ and $c_s$, the coupling parameter $\xi\lesssim 4$.

%Conclusion should be written by Ziwei 1.2.3

We only discuss the magnitude of the equilateral non-Gaussianity due to the peak contribution is on the equilateral configuration. It is imperative to compute the total shape function of axion inflation when considering the presence of heavy fields. The distinctive sharpness of the non-Gaussianity can, in principle, be distinguished from that observed in single axion inflation. Upon identifying specific models, the effects of heavy fields can potentially be detected in cosmological colliders in the future \cite{Chen:2009we,Baumann:2011nk,Noumi:2012vr,Arkani-Hamed:2015bza,Chen:2009zp,Assassi:2012zq,Sefusatti:2012ye,Norena:2012yi,Emami:2013lma,Liu:2015tza,Dimastrogiovanni:2015pla,Schmidt:2015xka,Chen:2015lza,Bonga:2015urq,Delacretaz:2015edn,Flauger:2016idt,Lee:2016vti,Delacretaz:2016nhw,Meerburg:2016zdz,Chen:2016uwp,Chen:2016hrz,An:2017hlx,Tong:2017iat,Iyer:2017qzw,An:2017rwo,Kumar:2017ecc,Riquelme:2017bxt,Saito:2018omt,Cabass:2018roz,Dimastrogiovanni:2018uqy,Bordin:2018pca,Arkani-Hamed:2018kmz,Kumar:2018jxz,Goon:2018fyu,Wu:2018lmx,Chua:2018dqh,Wang:2018tbf,McAneny:2019epy,Li:2019ves,Kim:2019wjo,Sleight:2019mgd,Biagetti:2019bnp,Sleight:2019hfp,Welling:2019bib,Alexander:2019vtb,Lu:2019tjj,Hook:2019zxa,Hook:2019vcn,ScheihingHitschfeld:2019tzr,Baumann:2019oyu,Wang:2019gbi,Liu:2019fag,Wang:2019gok,Wang:2020uic,Jazayeri:2023kji,Tong:2021wai,Pimentel:2022fsc,Wang:2022eop,Werth:2023pfl,Jazayeri:2023xcj}. One particularly intriguing model is rapid-turn inflation, characterized by an exceptionally rapid turning rate of the classical trajectory, resulting in a reduction of the speed of sound \eqref{cs} ($c_s^2<0$) . In this case, the light field would experience instability before Hubble crossing \cite{Garcia-Saenz:2018ifx,Fumagalli:2019noh,Garcia-Saenz:2018vqf}. In this paper, we only consider the massless gauge fields. It's also interesting to investigate the massive one, which also leads to rich phenomenology \cite{Liu:2015tza,Niu:2022quw,Niu:2022fki}. We leave all these discussions for future work.

In addition to the phenomenology, there are also several theoretical concerns to be addressed. Although during inflation, the strong backreaction can be ignored in the interesting parameter space of $\xi$, this parameter become inevitable large at the end of inflation, where the slow-roll is violated $\dot{\phi}/H\gtrsim 1$. Then the backreaction from EM fields to inflation becomes significant and change the power spectrum at late time \cite{Deskins:2013dwa,Adshead:2015pva,Cuissa:2018oiw}. When this parameter $\xi$ is large enough during inflation, one also need to concern the total inhomogeneous equation of motion during inflation \cite{Domcke:2020zez,Caravano:2022epk,Figueroa:2023oxc}. The heavy fields are also expected to impact the dynamics of scalar fields and EM fields. In this paper, we found in some regions of parameter space, the power spectrum of inverse decay is so negatively large that the total power spectrum may become negative. This could be the results of strongly coupling in the theories. There are two relevant parameters to this problem, $\xi$ and $\lambda$ in our discussion. Therefore we should also carefully calculate all of the contributions of inverse decay under perturbative control \cite{Ferreira:2015omg,Peloso:2016gqs}. These questions are also left for future investigations.

\acknowledgments
We would like to thank Zhengjun Mao for discussion in the early stage of this work and wish him well in his career. We also thank Ligong Bian, Shiyun Lu, Dong-Gang Wang, Yi Wang for helpful discussions. C-B.\,C. is supported by Japanese Government (MEXT) Scholarship and China Scholarship Council (CSC).
ZW is supported by JSPS short-term Postdoc Fellowship during his stay in Kobe University. SZ is supported by JSPS KAKENHI Grant Number 21F21026 during her stay in Kobe University and Natural Science Foundation of China under Grant No.12147102 during her stay in Chongqing University. SZ would also thank Jing Ren and Pengyuan Gao for hospitality when she visit IHEP where part of this work is finished. SZ would also like to thank 2023UFITS summer school where she got a chance to present this work and found several future collaborations.

\appendix

\section{Backreaction effects}
\label{appendix:be}
We have known one of the EM fields mode has tachyonic growth before crossing the horizon. We should study the effects of backreaction of EM fields to the inflation. If we neglect the no-growing mode $A_{-}$, the expectation values of two terms in the equations (\ref{ceom}) and (\ref{phieom}) are given by
\begin{align}
    \frac{1}{2}\langle\boldsymbol{E}^2+\boldsymbol{B}^2\rangle&=\frac{1}{4\pi^2a^4}\int{dk}k^2\left(|A_{+}'|^2+k^2|A_{+}|^2\right),\\
    \langle\boldsymbol{E}\cdot\boldsymbol{B}\rangle&=\frac{1}{4\pi^2a^4}\int{dk}k^3\frac{d}{d\tau}|A_{+}|^2.
\end{align}
The integration should include all the modes we are interested by introducing a UV cutoff $k_c\tau=-2\xi $. However, one can use the approximation solution (\ref{Asol}) in the integration interval $(8\xi )^{-1}\lesssim-k\tau\lesssim 2\xi $ to calculate. We can extend the integration to interval from $0$ to $\infty$ because most of contributions to the integration are in $(8\xi )^{-1}\lesssim-k\tau\lesssim 2\xi $. The contributions outer this interval is small and can be ignored \cite{Anber:2006xt,Durrer:2010mq}. Under this approximation, one have \cite{Barnaby:2011qe}
\begin{align}\label{br}
     \frac{1}{2}\langle\boldsymbol{E}^2+\boldsymbol{B}^2\rangle&\simeq 2.4\times 10^{-4}\frac{H^4}{\xi ^3}e^{2\pi\xi },\\ \langle\boldsymbol{E}\cdot\boldsymbol{B}\rangle&\simeq 1.4\times 10^{-4}\frac{H^4}{\xi ^4}e^{2\pi\xi }.
\end{align}

Firstly, to make sure the homogeneous inflationary background and the inflation is driven by the rolling of scalar field $\phi$: $3M_{\text{pl}}^2H^2\simeq V$. The energy of EM fields should be sub-dominated during inflation compared to the one of inflation, i.e., $\langle\boldsymbol{E}^2+\boldsymbol{B}^2\rangle/2\ll 3M_{\text{pl}}^2H^2$. Under (\ref{br}) this constrain can be written as
\begin{equation}
    \frac{H}{M_{\text{pl}}}\ll 146\xi ^{3/2}e^{-\pi\xi }.
\end{equation}
Secondly, the interaction of scalar $\phi$ and EM fields in the evolution equation of $\phi$ should also be small enough. If we assume that the terms of additional fields at least have the same order as the kinetic energy of axion field. We should have $(\alpha/f)\boldsymbol{E}\cdot\boldsymbol{B}\ll 3H\dot{\phi}$. Then we have another constrain
\begin{equation}\label{phibe}
    \frac{H^2}{|\dot{\phi}|}\ll 79\xi ^{3/2}e^{-\pi\xi }.
\end{equation}
These two constrains ensure the slow-roll and homogeneous inflationary background.

\section{Quantization in non-linear dispersion regime}
\label{ad1}
The quadratic Lagrangian of EFT at low-energy $\omega^2\ll M^2+p^2$ is 
\begin{equation}
    \mathcal{L}_{\zeta\zeta}=a^2M_{\text{pl}}^2\epsilon\left[\left(\frac{M^2c_s^{-2}-\nabla^2/a^2}{M^2-\nabla^2/a^2}\right)\zeta'^2-(\nabla \mathcal{\zeta})^2\right],
\end{equation}
where ``prime'' denotes the conformal time derivative. The conjugate momentum of $\zeta$ is
\begin{equation}
    p_{\zeta}=\frac{\partial\mathcal{L}_{\zeta\zeta}}{\partial \zeta'}=2a^2M_{\text{pl}}^2\epsilon\left(\frac{M^2c_s^{-2}-\nabla^2/a^2}{M^2-\nabla^2/a^2}\right)\zeta'.
\end{equation}
The equal-time canonical commutation relation $[\hat{\zeta}(\boldsymbol{x},\tau),\hat{p}_{\zeta}(\boldsymbol{x}',\tau)]=i\delta(\boldsymbol{x}-\boldsymbol{{x}'})$ implies
\begin{align}\label{cr}
    \zeta_k\zeta_k^{*'}-\zeta_k'\zeta_k^*=\frac{i}{2a^2M_{\text{pl}}^2\epsilon}\left(\frac{M^2+p^2}{M^2c_s^{-2}+p^2}\right).
\end{align}
Now we can solve $\zeta$ by writing down the equation of motion in the limit $M^2\ll p^2\ll M^2c_s^{-2}$ and de Sitter limit,
\begin{equation}
    \zeta_k''-\frac{4}{\tau}\zeta_k'+\frac{k^4H^2\tau^2}{\Lambda_{\text{UV}}}\zeta_k=0.
\end{equation}
This equation has general analytical solution
\begin{equation}
    \zeta_k=(-\tau)^{5/2}\left[c_1H^{(1)}_{5/4}(u)+c_2H^{(2)}_{5/4}(u)\right], \ \ \ \ \ \ \ u\equiv\frac{H}{2\Lambda_{\text{UV}}}k^2\tau^2.
\end{equation}
We choose the positive frequency solution initially by fixing $c_2=0$. To impose the commutation relation (\ref{cr}) we use the derivative of the Hankel function of the first kind
\begin{equation}\label{dH1}
    \frac{d}{dz}H_{\nu}^{(1)}(z)=H_{\nu-1}^{(1)}(z)-\frac{\nu H_{\nu}^{(1)}(z)}{z},
\end{equation}
we have 
\begin{equation}
    \zeta_k'=-2c_1(-\tau)^{3/2}uH^{(1)}_{1/4}(u).
\end{equation}
Expanding the Hankel function of the first kind near infinity
\begin{equation}
    H^{(1)}_{\nu}(z)=\sqrt{\frac{\pi}{2z}}e^{iz-i\pi(-3+2\nu)/4}+\cdots,\ \ \ \ \ \ \ \ z\to\infty.
\end{equation}
and imposing the commutation relation on sub-horizon, we can obtain the coefficient $c_1$ as
\begin{equation}
    |c_1|=\sqrt{\frac{\pi}{8}}\frac{H^2}{\sqrt{2\epsilon}M_{\text{pl}}}\frac{k}{\Lambda_{\text{UV}}}.
\end{equation}
Another useful identity is the expanding of the Hankel function of the first kind near $z=0$
\begin{equation}\label{hankel0}
    H^{(1)}_{\nu}(z)=i\frac{2^{\nu}\Gamma[\nu]}{\pi z^{\nu}}+\frac{z^{\nu}}{2^{\nu}\Gamma[1+\nu]}+\cdots,\ \ \ \ \ \ \ \ z\to0,
\end{equation}
where we only show the leading order terms of the real parts and the imaginary parts. Then we can calculate the power spectrum at late time and some integrals in appendix \ref{sec:estimation}.

\section{Estimation of power spectrum and bispectrum}
\label{sec:estimation}
We estimate some integrals of the power spectrum and bispectrum of the main text in this appendix. To compute the two-point and three-point correlation functions, we need identities
\begin{align}\label{id}
    &\boldsymbol{\nabla}\times[\boldsymbol{\epsilon}_{\pm}(\boldsymbol{k})e^{i\boldsymbol{k}\cdot\boldsymbol{x}}]=i\boldsymbol{k}\times\boldsymbol{\epsilon}_{\pm}(\boldsymbol{k})=\pm k\boldsymbol{\epsilon}_{\pm}(\boldsymbol{k}),\nonumber\\
    &\left|\boldsymbol{\epsilon}_{\pm}(\boldsymbol{q})\cdot\boldsymbol{\epsilon}_{\pm}(\boldsymbol{k}-\boldsymbol{q})\right|=\frac{1}{2}\left|1+\frac{|\boldsymbol{q}|^2-\boldsymbol{q}\cdot\boldsymbol{k}}{|\boldsymbol{q}-\boldsymbol{k}||\boldsymbol{q}|}\right|.
\end{align}
and de-Sitter approxiamtion $aH=-1/\tau$ to obtain the correlation of sources  $J_{\boldsymbol{k}}$ \cite{Barnaby:2011vw}
\begin{align}
    \langle J_1J_2\rangle=
    &\frac{\xi^2H^2\tau_1\tau_2e^{4\pi\xi }}{16\epsilon M_{\text{pl}}^2}\delta(\boldsymbol{k}_1+\boldsymbol{k}_2)\int d^3q\left[1+\frac{|\boldsymbol{q}|^2-\boldsymbol{q}\cdot\boldsymbol{k}_1}{|\boldsymbol{q}-\boldsymbol{k}_1||\boldsymbol{q}|}\right]^2|\boldsymbol{q}|^{1/2}|\boldsymbol{q}-\boldsymbol{k}_1|^{1/2}\nonumber\\
    &\times \left(|\boldsymbol{q}|^{1/2}+|\boldsymbol{q}-\boldsymbol{k}_1|^{1/2}\right)^2e^{-2\sqrt{2\xi }(\sqrt{-\tau_1}+\sqrt{-\tau_2})(\sqrt{|\boldsymbol{q}|}+\sqrt{|\boldsymbol{q}-\boldsymbol{k}_1}|)}.\label{JJf}\\
    \langle J_1J_2J_3\rangle=&\frac{\xi^3H^3e^{6\pi\xi }\tau_1\tau_2\tau_3}{(2\epsilon)^{3/2}M_{\text{pl}}^3}\delta (\boldsymbol{k}_1+\boldsymbol{k}_2+\boldsymbol{k}_3)\int d^3q\ |\boldsymbol{q}|^{1/2}|\boldsymbol{k}_1-\boldsymbol{q}|^{1/2}|\boldsymbol{k}_3+\boldsymbol{q}|^{1/2}\nonumber\\
    &\times\left[\boldsymbol{\epsilon}(\boldsymbol{q})\cdot\boldsymbol{\epsilon}(\boldsymbol{k}_1-\boldsymbol{q})\right]\left(|\boldsymbol{q}|^{1/2}+|\boldsymbol{k}_1-\boldsymbol{q}|^{1/2}\right)e^{-2\sqrt{-2\xi \tau_1}(\sqrt{|\boldsymbol{q}|}+\sqrt{|\boldsymbol{k}_1-\boldsymbol{q}}|)}\nonumber\\
    &\times\left[\boldsymbol{\epsilon}(\boldsymbol{q}-\boldsymbol{k}_1)\cdot\boldsymbol{\epsilon}(-\boldsymbol{q}-\boldsymbol{k}_3)\right]\left(|\boldsymbol{k}_1-\boldsymbol{q}|^{1/2}+|\boldsymbol{k}_3+\boldsymbol{q}|^{1/2}\right)\nonumber\\
    &\ \ \ \ e^{-2\sqrt{-2\xi \tau_2}(\sqrt{|\boldsymbol{k}_1-\boldsymbol{q}|}+\sqrt{|\boldsymbol{k}_3+\boldsymbol{q}}|)}\nonumber\\
    &\times\left[\boldsymbol{\epsilon}(\boldsymbol{q}+\boldsymbol{k}_3)\cdot\boldsymbol{\epsilon}(-\boldsymbol{q})\right]\left(|\boldsymbol{k}_3+\boldsymbol{q}|^{1/2}+|\boldsymbol{q}|^{1/2}\right)e^{-2\sqrt{-2\xi \tau_1}(\sqrt{|\boldsymbol{k}_3+\boldsymbol{q}|}+\sqrt{|\boldsymbol{q}}|)}.\label{JJJf}
\end{align}
Here we have used the approximation solution of gauge potential (\ref{Asol}). Because the impacts of heavy fields is on the free propagation of the curvature fluctuation, it's concise to define functions 
\begin{align}
    \mathbb{F}_2[\xi;\mathcal{I}_1,\mathcal{I}_2]:=&\int d^3q_*\left[1-\frac{\boldsymbol{q}_*\cdot\boldsymbol{Q}_1}{|\boldsymbol{Q}_1||\boldsymbol{q}_*|}\right]^2|\boldsymbol{q}_*|^{1/2}|\boldsymbol{Q}_1|^{1/2}\left(|\boldsymbol{q}_*|^{1/2}+|\boldsymbol{Q}_1|^{1/2}\right)^2 \nonumber\\
     &\times\mathcal{I}_1\left[2\sqrt{2\xi }\left(|\boldsymbol{q}_*|^{1/2}+|\boldsymbol{Q}_1|^{1/2}\right)\right]\mathcal{I}_2\left[2\sqrt{2\xi }\left(|\boldsymbol{q}_*|^{1/2}+|\boldsymbol{Q}_1|^{1/2}\right)\right],\\
    \mathbb{F}_3[\xi;\mathcal{I}_1,\mathcal{I}_2,\mathcal{I}_3]:=&\int d^3\boldsymbol{q}_*\ |\boldsymbol{q}_*|^{1/2}|\boldsymbol{Q}_1|^{1/2}|\boldsymbol{q}_*+x_3\hat{k}_3|^{1/2}\nonumber\\
    &\times\left[\boldsymbol{\epsilon}(\boldsymbol{q}_*)\cdot\boldsymbol{\epsilon}(\boldsymbol{Q}_1)\right]\left(|\boldsymbol{q}_*|^{1/2}+|\boldsymbol{Q}_1|^{1/2}\right)\mathcal{I}_1\left[2\sqrt{2\xi }\left(|\boldsymbol{q}_*|^{1/2}+|\boldsymbol{Q}_1|^{1/2}\right)\right]\nonumber\\
    &\times\left[\boldsymbol{\epsilon}(\boldsymbol{Q}_1)\cdot\boldsymbol{\epsilon}(-\boldsymbol{Q}_3(x_3))\right]\left(|\boldsymbol{Q}_1|^{1/2}+|\boldsymbol{Q}_3(x_3)|^{1/2}\right)\nonumber\\
    &\ \ \ \ \mathcal{I}_2\left[2\sqrt{2\xi }\left(|\boldsymbol{Q}_1|^{1/2}+|\boldsymbol{Q}_3(x_3)|^{1/2}\right)\right]\nonumber\\ 
    &\times\left[\boldsymbol{\epsilon}(\boldsymbol{Q}_3(x_3))\cdot\boldsymbol{\epsilon}(-\boldsymbol{q}_*)\right]\left(|\boldsymbol{q}_*|^{1/2}+|\boldsymbol{Q}_3(x_3)|^{1/2}\right)\nonumber\\
    &\ \ \ \ \mathcal{I}_3\left[2\sqrt{2\xi }\left(|\boldsymbol{Q}_3(x_3)|^{1/2}+|\boldsymbol{q}_*|^{1/2}\right)\right],
\end{align}
where we have defined the dimensionaless integration variable $\boldsymbol{q}_*\equiv\boldsymbol{q}/|\boldsymbol{k}_1|$, $\boldsymbol{Q}_1\equiv \hat{k}_1-\boldsymbol{q}_*$ and $\boldsymbol{Q}_3(x_3)\equiv \hat{k}_3+\boldsymbol{q}_*$. 

In the linear dispersion regime,  the total contributions of the inverse decay processess to the power spectrum is given by (\ref{twopf}), where the dimensionaless functions $f_2^{GG}(\xi)$, $f_2^{GF}(\xi)$ and $f_2^{FF}(\xi)$ are evaluated by
\begin{align}
    f_2^{GG}(\xi)&=\frac{\xi^2}{8\pi}\mathbb{F}_2[\xi,\mathcal{I}_{G},\mathcal{I}_{G}],\\ 
    f_2^{GF}(\xi)&=\frac{\xi^2}{8\pi}\mathbb{F}_2[\xi,\mathcal{I}_{G},\mathcal{I}_{F}],\\
    f_2^{FF}(\xi)&=\frac{\xi^2}{8\pi}\mathbb{F}_2[\xi,\mathcal{I}_{F},\mathcal{I}_{F}], 
\end{align}
and the functions $\mathcal{I}_{G}$ and $\mathcal{I}_{F}$ are given by
\begin{align}
    \mathcal{I}_{G}(z)&\equiv\int_{0}^{\infty}dx\ \left[\sin{(c_sx)}-c_sx\cos{(c_sx)}\right]e^{-z\sqrt{x}},\label{f21}\\
    \mathcal{I}_{F}(z)&\equiv-\int_{0}^{\infty}dx\ c_s^2x^2\sin{(c_sx)}e^{-z\sqrt{x}}.\label{f22}
\end{align}
After rescaling the integral variable $x_c=c_sx$, which is equivalent to rescaling the spatial coordinates $\boldsymbol{x}\to c_s^{-1}\boldsymbol{x}$ and parameter $\xi\to \xi_c\equiv c_s^{-1}\xi$. Then we can directly adopt the results from \cite{Barnaby:2011vw} for $f_2^{GG}$, which have been shown in (\ref{f2A}). For small $x_c$, $x_c^2\sin{x_c}\simeq x_c^3$ hence $\mathcal{I}_{F}\simeq -3\mathcal{I}_{F}$. Then functions $f_2^{GF}$ and $f_{2}^{FF}$ can be estimated by $-6f_2^{GF}$ and $9f_2^{GG}$ respectively.

\begin{figure}[tbp]
\centering
\includegraphics[scale=0.8]{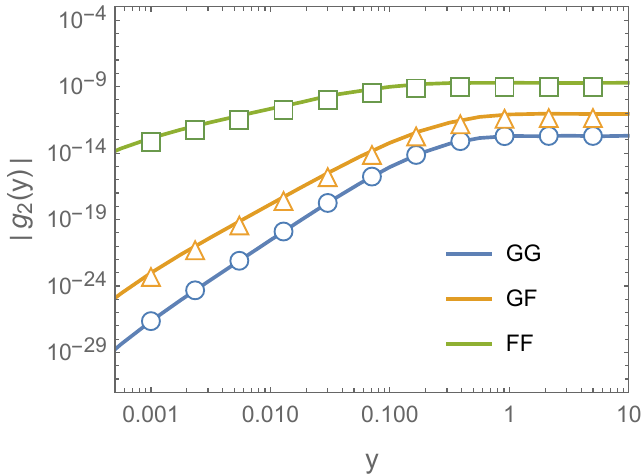}
\caption{\label{fig:g2tot}$\Lambda_{\text{UV}}/H=100$ and $\xi=1$. The integral (\ref{g2GG})-(\ref{g2FF}) as functions of upper limit of time integration $y$. The curves are  approximations (\ref{JA})-(\ref{JB}) while the discrete points are exact calculations (\ref{JAe})-(\ref{JBe}).}
\end{figure}

In the non-linear dispersion regime, the contributions of inverse decay processes is (\ref{twopf2}), where the dimensionless functions $g_2$ are
\begin{align}
    g_2^{GG}(\xi)&=\frac{2^{-11/2}\pi\xi_{\text{UV}}^2}{ \Gamma^2(5/4)}\mathbb{F}_2[\xi_{\text{UV}},\mathcal{J}_{G},\mathcal{J}_{G}],\label{g2GG}\\
    g_2^{GF}(\xi)&=\frac{2^{-13/2}\pi\xi_{\text{UV}}^2}{ \Gamma^2(5/4)}\mathbb{F}_2[\xi_{\text{UV}},\mathcal{J}_{G},\mathcal{J}_{F}],\label{g2GF}\\
    g_2^{FF}(\xi)&=\frac{2^{-15/2}\pi\xi_{\text{UV}}^2}{ \Gamma^2(5/4)}\mathbb{F}_2[\xi_{\text{UV}},\mathcal{J}_{F},\mathcal{J}_{F}]\label{g2FF},
\end{align}
where $\xi_{\text{UV}}\equiv\xi\sqrt{2\Lambda_{\text{UV}}/H}$ and
\begin{align}
    \mathcal{J}_{G}(z)&=\int_{0}^{\infty}dy\ y^{5/2}\text{Re}\left[H^{(1)}_{5/4}(y^2)\right]e^{-z\sqrt{y}},\label{JAe}\\
    \mathcal{J}_{F}(z)&=-2\int_0^{\infty}dy\ y^{5/2}\text{Re}\left[H_{1/4}^{(1)}(y^2)\right]e^{-z\sqrt{y}}\label{JBe}.
\end{align}[eq].
By using the zero limit of the Hankel function (\ref{hankel0}), the integral can also be simplified when the argument of $\mathcal{J}$ is much greater than one,
\begin{align}
    \mathcal{J}_{G}(z)\simeq&\int_0^{\infty}dy\ \frac{y^5}{2^{5/4}\Gamma(9/4)}e^{-z\sqrt{y}}\simeq\frac{2.96\times10^7}{z^{12}}, \ \ \ \ \ \ \ z\gg 1,\label{JA}\\
    \mathcal{J}_{F}(z)\simeq& -\int_{0}^{\infty}dy\ \frac{y^3}{2^{-3/4}\Gamma(5/4)}e^{-z\sqrt{y}}\simeq -\frac{1.87\times 10^4}{z^8}, \ \ \ \ \ \ \ z\gg 1\label{JB}.
\end{align}
For $z\sim 2\sqrt{2\xi_{\text{UV}}}|\boldsymbol{q}_{*}|^{1/2}$ in the integrals (\ref{g2GG})-(\ref{g2FF}), the peaks are at $|\boldsymbol{q}_*|\sim\mathcal{O}(1)$. Hence these approxiamtions are good enough for interesting regime $\xi\gtrsim1$ because $\Lambda_{\text{UV}}/H$ is large. We show in Figure \ref{fig:g2tot} the agreements of these approximations as functions of upper limit $y$ of the time integration, by choosing $\Lambda_{\text{UV}}=100H$ and a not so large $\xi=1$.

Similarly for the three-point correlation function (\ref{bis1}) and (\ref{bis2}), but we also need to calculate the products of polarization operators in the integrals. For any wavenumber $\boldsymbol{k}=(A,B,C)$, the corresponding polarization operators are
\begin{align}
    \boldsymbol{\epsilon}_{\pm}(\boldsymbol{k})=\frac{1}{N}\left(\frac{C}{k}\frac{A}{k_{xy}}\mp i\frac{B}{k_{xy}},\frac{C}{k}\frac{B}{k_{xy}}\pm i\frac{A}{k_{xy}},-\frac{k_{xy}}{k}\right),
\end{align}
where $N$ is the normal factor, $k=\sqrt{A^2+B^2+C^2}$ and $k_{xy}=\sqrt{A^2+B^2}$. We can also use $\mathcal{I}$ and $\mathcal{J}$. Then in the linear dispersion regime we have
\begin{align}
    f_3^{GGG}(\xi;x_2,x_3)&=\frac{5}{3\pi}\frac{\xi^3}{x_2x_3(1+x_2^3+x_3^3)}\mathbb{F}_3[\xi,\mathcal{I}_{G},\mathcal{I}_{G},\mathcal{I}_{G}],\label{f3A}\\
    f_3^{GGF}(\xi;x_2,x_3)&=\frac{5}{3\pi}\frac{\xi^3}{x_2x_3(1+x_2^3+x_3^3)}\mathbb{F}_3[\xi,\mathcal{I}_{G},\mathcal{I}_{G},\mathcal{I}_{F}],\\
    f_3^{GFF}(\xi;x_2,x_3)&=\frac{5}{3\pi}\frac{\xi^3}{x_2x_3(1+x_2^3+x_3^3)}\mathbb{F}_3[\xi,\mathcal{I}_{G},\mathcal{I}_{F},\mathcal{I}_{F}],\\
    f_3^{FFF}(\xi;x_2,x_3)&=\frac{5}{3\pi}\frac{\xi^3}{x_2x_3(1+x_2^3+x_3^3)}\mathbb{F}_3[\xi,\mathcal{I}_{F},\mathcal{I}_{F},\mathcal{I}_{F}]\label{f3E}.
\end{align}
While in the non-linear dispersion limit,
\begin{align}
    g_3^{GGG}(\xi;x_2,x_3)&=\frac{2^{-3/4}\cdot5\pi^2}{3\Gamma^3[5/4]}\frac{\xi_{\text{UV}}^3}{x_2x_3(1+x_2^3+x_3^3)}\mathbb{F}_3[\xi_{\text{UV}},\mathcal{J}_{G},\mathcal{J}_{G},\mathcal{J}_{G}],\label{g3A}\\
    g_3^{GGF}(\xi;x_2,x_3)&=\frac{2^{-7/4}\cdot5\pi^2}{3\Gamma^3[5/4]}\frac{\xi_{\text{UV}}^3}{x_2x_3(1+x_2^3+x_3^3)}\mathbb{F}_3[\xi_{\text{UV}},\mathcal{J}_{G},\mathcal{J}_{G},\mathcal{J}_{F}],\\
    g_3^{GFF}(\xi;x_2,x_3)&=\frac{2^{-11/4}\cdot5\pi^2}{3\Gamma^3[5/4]}\frac{\xi_{\text{UV}}^3}{x_2x_3(1+x_2^3+x_3^3)}\mathbb{F}_3[\xi_{\text{UV}},\mathcal{J}_{G},\mathcal{J}_{F},\mathcal{J}_{F}],\\
    g_3^{FFF}(\xi;x_2,x_3)&=\frac{2^{-15/4}\cdot5\pi^2}{3\Gamma^3[5/4]}\frac{\xi_{\text{UV}}^3}{x_2x_3(1+x_2^3+x_3^3)}\mathbb{F}_3[\xi_{\text{UV}},\mathcal{J}_{F},\mathcal{J}_{F},\mathcal{J}_{F}].\label{g3E}
\end{align}



\begin{thebibliography}{99}

\bibitem{Weinberg:1975ui}
S.~Weinberg,
``The U(1) Problem,''
Phys. Rev. D \textbf{11}, 3583-3593 (1975)
doi:10.1103/PhysRevD.11.3583

\bibitem{tHooft:1976rip}
G.~'t Hooft,
``Symmetry Breaking Through Bell-Jackiw Anomalies,''
Phys. Rev. Lett. \textbf{37}, 8-11 (1976)
doi:10.1103/PhysRevLett.37.

\bibitem{tHooft:1976snw}
G.~'t Hooft,
``Computation of the Quantum Effects Due to a Four-Dimensional Pseudoparticle,''
Phys. Rev. D \textbf{14}, 3432-3450 (1976)
[erratum: Phys. Rev. D \textbf{18}, 2199 (1978)]
doi:10.1103/PhysRevD.14.3432

\bibitem{Adler:1969gk}
S.~L.~Adler,
``Axial vector vertex in spinor electrodynamics,''
Phys. Rev. \textbf{177}, 2426-2438 (1969)
doi:10.1103/PhysRev.177.2426

\bibitem{Bell:1969ts}
J.~S.~Bell and R.~Jackiw,
``A PCAC puzzle: $\pi^0 \to \gamma \gamma$ in the $\sigma$ model,''
Nuovo Cim. A \textbf{60}, 47-61 (1969)
doi:10.1007/BF02823296

\bibitem{Bardeen:1969md}
W.~A.~Bardeen,
``Anomalous Ward identities in spinor field theories,''
Phys. Rev. \textbf{184}, 1848-1857 (1969)
doi:10.1103/PhysRev.184.1848

\bibitem{Arvanitaki:2009fg}
A.~Arvanitaki, S.~Dimopoulos, S.~Dubovsky, N.~Kaloper and J.~March-Russell,
``String Axiverse,''
Phys. Rev. D \textbf{81}, 123530 (2010)
doi:10.1103/PhysRevD.81.123530
[arXiv:0905.4720 [hep-th]].

\bibitem{Marsh:2015xka}
D.~J.~E.~Marsh,
``Axion Cosmology,''
Phys. Rept. \textbf{643}, 1-79 (2016)
doi:10.1016/j.physrep.2016.06.005
[arXiv:1510.07633 [astro-ph.CO]].

\bibitem{Hui:2016ltb}
L.~Hui, J.~P.~Ostriker, S.~Tremaine and E.~Witten,
``Ultralight scalars as cosmological dark matter,''
Phys. Rev. D \textbf{95}, no.4, 043541 (2017)
doi:10.1103/PhysRevD.95.043541
[arXiv:1610.08297 [astro-ph.CO]].

\bibitem{Niemeyer:2019aqm}
J.~C.~Niemeyer,
``Small-scale structure of fuzzy and axion-like dark matter,''
doi:10.1016/j.ppnp.2020.103787
[arXiv:1912.07064 [astro-ph.CO]].

\bibitem{Freese:1990rb}
K.~Freese, J.~A.~Frieman and A.~V.~Olinto,
``Natural inflation with pseudo - Nambu-Goldstone bosons,''
Phys. Rev. Lett. \textbf{65}, 3233-3236 (1990)
doi:10.1103/PhysRevLett.65.3233

\bibitem{Adams:1992bn}
F.~C.~Adams, J.~R.~Bond, K.~Freese, J.~A.~Frieman and A.~V.~Olinto,
``Natural inflation: Particle physics models, power law spectra for large scale structure, and constraints from COBE,''
Phys. Rev. D \textbf{47}, 426-455 (1993)
doi:10.1103/PhysRevD.47.426
[arXiv:hep-ph/9207245 [hep-ph]].

\bibitem{Moroi:2000jr}
T.~Moroi and T.~Takahashi,
``Constraints on natural inflation from cosmic microwave background,''
Phys. Lett. B \textbf{503}, 376-383 (2001)
doi:10.1016/S0370-2693(01)00246-5
[arXiv:hep-ph/0010197 [hep-ph]].

\bibitem{Dimopoulos:2005ac}
S.~Dimopoulos, S.~Kachru, J.~McGreevy and J.~G.~Wacker,
``N-flation,''
JCAP \textbf{08}, 003 (2008)
doi:10.1088/1475-7516/2008/08/003
[arXiv:hep-th/0507205 [hep-th]].

\bibitem{Planck:2018jri}
Y.~Akrami \textit{et al.} [Planck],
``Planck 2018 results. X. Constraints on inflation,''
Astron. Astrophys. \textbf{641}, A10 (2020)
doi:10.1051/0004-6361/201833887
[arXiv:1807.06211 [astro-ph.CO]].

\bibitem{Garretson:1992vt}
W.~D.~Garretson, G.~B.~Field and S.~M.~Carroll,
``Primordial magnetic fields from pseudoGoldstone bosons,''
Phys. Rev. D \textbf{46}, 5346-5351 (1992)
doi:10.1103/PhysRevD.46.5346
[arXiv:hep-ph/9209238 [hep-ph]]

\bibitem{Anber:2006xt}
M.~M.~Anber and L.~Sorbo,
``N-flationary magnetic fields,''
JCAP \textbf{10}, 018 (2006)
doi:10.1088/1475-7516/2006/10/018
[arXiv:astro-ph/0606534 [astro-ph]].

\bibitem{Durrer:2010mq}
R.~Durrer, L.~Hollenstein and R.~K.~Jain,
``Can slow roll inflation induce relevant helical magnetic fields?,''
JCAP \textbf{03}, 037 (2011)
doi:10.1088/1475-7516/2011/03/037
[arXiv:1005.5322 [astro-ph.CO]].

\bibitem{Barnaby:2010vf}
N.~Barnaby and M.~Peloso,
``Large Nongaussianity in Axion Inflation,''
Phys. Rev. Lett. \textbf{106}, 181301 (2011)
doi:10.1103/PhysRevLett.106.181301
[arXiv:1011.1500 [hep-ph]].

\bibitem{Barnaby:2011vw}
N.~Barnaby, R.~Namba and M.~Peloso,
``Phenomenology of a Pseudo-Scalar Inflaton: Naturally Large Nongaussianity,''
JCAP \textbf{04}, 009 (2011)
doi:10.1088/1475-7516/2011/04/009
[arXiv:1102.4333 [astro-ph.CO]].

\bibitem{Barnaby:2011qe}
N.~Barnaby, E.~Pajer and M.~Peloso,
``Gauge Field Production in Axion Inflation: Consequences for Monodromy, non-Gaussianity in the CMB, and Gravitational Waves at Interferometers,''
Phys. Rev. D \textbf{85}, 023525 (2012)
doi:10.1103/PhysRevD.85.023525
[arXiv:1110.3327 [astro-ph.CO]].

\bibitem{Braden:2010wd}
J.~Braden, L.~Kofman and N.~Barnaby,
``Reheating the Universe After Multi-Field Inflation,''
JCAP \textbf{07}, 016 (2010)
doi:10.1088/1475-7516/2010/07/016
[arXiv:1005.2196 [hep-th]].

\bibitem{Adshead:2015pva}
P.~Adshead, J.~T.~Giblin, T.~R.~Scully and E.~I.~Sfakianakis,
``Gauge-preheating and the end of axion inflation,''
JCAP \textbf{12}, 034 (2015)
doi:10.1088/1475-7516/2015/12/034
[arXiv:1502.06506 [astro-ph.CO]].

\bibitem{Adshead:2016iae}
P.~Adshead, J.~T.~Giblin, T.~R.~Scully and E.~I.~Sfakianakis,
``Magnetogenesis from axion inflation,''
JCAP \textbf{10}, 039 (2016)
doi:10.1088/1475-7516/2016/10/039
[arXiv:1606.08474 [astro-ph.CO]].

\bibitem{Cuissa:2018oiw}
J.~R.~C.~Cuissa and D.~G.~Figueroa,
``Lattice formulation of axion inflation. Application to preheating,''
JCAP \textbf{06}, 002 (2019)
doi:10.1088/1475-7516/2019/06/002
[arXiv:1812.03132 [astro-ph.CO]].

\bibitem{Sorbo:2011rz}
L.~Sorbo,
``Parity violation in the Cosmic Microwave Background from a pseudoscalar inflaton,''
JCAP \textbf{06}, 003 (2011)
doi:10.1088/1475-7516/2011/06/003
[arXiv:1101.1525 [astro-ph.CO]].

\bibitem{Cook:2011hg}
J.~L.~Cook and L.~Sorbo,
``Particle production during inflation and gravitational waves detectable by ground-based interferometers,''
Phys. Rev. D \textbf{85}, 023534 (2012)
[erratum: Phys. Rev. D \textbf{86}, 069901 (2012)]
doi:10.1103/PhysRevD.85.023534
[arXiv:1109.0022 [astro-ph.CO]].

\bibitem{Anber:2012du}
M.~M.~Anber and L.~Sorbo,
``Non-Gaussianities and chiral gravitational waves in natural steep inflation,''
Phys. Rev. D \textbf{85}, 123537 (2012)
doi:10.1103/PhysRevD.85.123537
[arXiv:1203.5849 [astro-ph.CO]].

\bibitem{Bartolo:2016ami}
N.~Bartolo, C.~Caprini, V.~Domcke, D.~G.~Figueroa, J.~Garcia-Bellido, M.~C.~Guzzetti, M.~Liguori, S.~Matarrese, M.~Peloso and A.~Petiteau, \textit{et al.}
``Science with the space-based interferometer LISA. IV: Probing inflation with gravitational waves,''
JCAP \textbf{12}, 026 (2016)
doi:10.1088/1475-7516/2016/12/026
[arXiv:1610.06481 [astro-ph.CO]].

\bibitem{Bastero-Gil:2022fme}
M.~Bastero-Gil and A.~T.~Manso,
``Parity violating gravitational waves at the end of inflation,''
JCAP \textbf{08}, 001 (2023)
doi:10.1088/1475-7516/2023/08/001
[arXiv:2209.15572 [gr-qc]].

\bibitem{Linde:2012bt}
A.~Linde, S.~Mooij and E.~Pajer,
``Gauge field production in supergravity inflation: Local non-Gaussianity and primordial black holes,''
Phys. Rev. D \textbf{87}, no.10, 103506 (2013)
doi:10.1103/PhysRevD.87.103506
[arXiv:1212.1693 [hep-th]].

\bibitem{Bugaev:2013fya}
E.~Bugaev and P.~Klimai,
``Axion inflation with gauge field production and primordial black holes,''
Phys. Rev. D \textbf{90}, no.10, 103501 (2014)
doi:10.1103/PhysRevD.90.103501
[arXiv:1312.7435 [astro-ph.CO]].

\bibitem{Garcia-Bellido:2016dkw}
J.~Garcia-Bellido, M.~Peloso and C.~Unal,
``Gravitational waves at interferometer scales and primordial black holes in axion inflation,''
JCAP \textbf{12}, 031 (2016)
doi:10.1088/1475-7516/2016/12/031
[arXiv:1610.03763 [astro-ph.CO]].

\bibitem{Domcke:2017fix}
V.~Domcke, F.~Muia, M.~Pieroni and L.~T.~Witkowski,
``PBH dark matter from axion inflation,''
JCAP \textbf{07}, 048 (2017)
doi:10.1088/1475-7516/2017/07/048
[arXiv:1704.03464 [astro-ph.CO]].

\bibitem{Caprini:2014mja}
C.~Caprini and L.~Sorbo,
``Adding helicity to inflationary magnetogenesis,''
JCAP \textbf{10}, 056 (2014)
doi:10.1088/1475-7516/2014/10/056
[arXiv:1407.2809 [astro-ph.CO]].

\bibitem{Fujita:2015iga}
T.~Fujita, R.~Namba, Y.~Tada, N.~Takeda and H.~Tashiro,
``Consistent generation of magnetic fields in axion inflation models,''
JCAP \textbf{05}, 054 (2015)
doi:10.1088/1475-7516/2015/05/054
[arXiv:1503.05802 [astro-ph.CO]].

\bibitem{Patel:2019isj}
T.~Patel, H.~Tashiro and Y.~Urakawa,
``Resonant magnetogenesis from axions,''
JCAP \textbf{01}, 043 (2020)
doi:10.1088/1475-7516/2020/01/043
[arXiv:1909.00288 [astro-ph.CO]].

\bibitem{Planck:2019kim}
Y.~Akrami \textit{et al.} [Planck],
``Planck 2018 results. IX. Constraints on primordial non-Gaussianity,''
Astron. Astrophys. \textbf{641}, A9 (2020)
doi:10.1051/0004-6361/201935891
[arXiv:1905.05697 [astro-ph.CO]].

\bibitem{Cheung:2007st}
C.~Cheung, P.~Creminelli, A.~L.~Fitzpatrick, J.~Kaplan and L.~Senatore,
``The Effective Field Theory of Inflation,''
JHEP \textbf{03}, 014 (2008)
doi:10.1088/1126-6708/2008/03/014
[arXiv:0709.0293 [hep-th]].

\bibitem{Weinberg:2008hq}
S.~Weinberg,
``Effective Field Theory for Inflation,''
Phys. Rev. D \textbf{77}, 123541 (2008)
doi:10.1103/PhysRevD.77.123541
[arXiv:0804.4291 [hep-th]].

\bibitem{Senatore:2010wk}
L.~Senatore and M.~Zaldarriaga,
``The Effective Field Theory of Multifield Inflation,''
JHEP \textbf{04}, 024 (2012)
doi:10.1007/JHEP04(2012)024
[arXiv:1009.2093 [hep-th]].

\bibitem{Achucarro:2010da}
A.~Achucarro, J.~O.~Gong, S.~Hardeman, G.~A.~Palma and S.~P.~Patil,
``Features of heavy physics in the CMB power spectrum,''
JCAP \textbf{01}, 030 (2011)
doi:10.1088/1475-7516/2011/01/030
[arXiv:1010.3693 [hep-ph]].

\bibitem{Achucarro:2010jv}
A.~Achucarro, J.~O.~Gong, S.~Hardeman, G.~A.~Palma and S.~P.~Patil,
``Mass hierarchies and non-decoupling in multi-scalar field dynamics,''
Phys. Rev. D \textbf{84}, 043502 (2011)
doi:10.1103/PhysRevD.84.043502
[arXiv:1005.3848 [hep-th]].

\bibitem{Achucarro:2012yr}
A.~Achucarro, V.~Atal, S.~Cespedes, J.~O.~Gong, G.~A.~Palma and S.~P.~Patil,
``Heavy fields, reduced speeds of sound and decoupling during inflation,''
Phys. Rev. D \textbf{86}, 121301 (2012)
doi:10.1103/PhysRevD.86.121301
[arXiv:1205.0710 [hep-th]].

\bibitem{Cespedes:2012hu}
S.~Cespedes, V.~Atal and G.~A.~Palma,
``On the importance of heavy fields during inflation,''
JCAP \textbf{05}, 008 (2012)
doi:10.1088/1475-7516/2012/05/008
[arXiv:1201.4848 [hep-th]].

\bibitem{Achucarro:2012sm}
A.~Achucarro, J.~O.~Gong, S.~Hardeman, G.~A.~Palma and S.~P.~Patil,
``Effective theories of single field inflation when heavy fields matter,''
JHEP \textbf{05}, 066 (2012)
doi:10.1007/JHEP05(2012)066
[arXiv:1201.6342 [hep-th]].

\bibitem{Baumann:2011su}
D.~Baumann and D.~Green,
``Equilateral Non-Gaussianity and New Physics on the Horizon,''
JCAP \textbf{09}, 014 (2011)
doi:10.1088/1475-7516/2011/09/014
[arXiv:1102.5343 [hep-th]].

\bibitem{Gwyn:2012mw}
R.~Gwyn, G.~A.~Palma, M.~Sakellariadou and S.~Sypsas,
``Effective field theory of weakly coupled inflationary models,''
JCAP \textbf{04}, 004 (2013)
doi:10.1088/1475-7516/2013/04/004
[arXiv:1210.3020 [hep-th]].

\bibitem{Ballardini:2019rqh}
M.~Ballardini, M.~Braglia, F.~Finelli, G.~Marozzi and A.~A.~Starobinsky,
``Energy-momentum tensor and helicity for gauge fields coupled to a pseudo-scalar inflaton,''
Phys. Rev. D \textbf{100}, no.12, 123542 (2019)
[erratum: Phys. Rev. D \textbf{105}, no.6, 069905 (2022)]
doi:10.1103/PhysRevD.100.123542
[arXiv:1910.13448 [gr-qc]].

\bibitem{Domcke:2019qmm}
V.~Domcke, Y.~Ema and K.~Mukaida,
``Chiral Anomaly, Schwinger Effect, Euler-Heisenberg Lagrangian, and application to axion inflation,''
JHEP \textbf{02}, 055 (2020)
doi:10.1007/JHEP02(2020)055
[arXiv:1910.01205 [hep-ph]].

\bibitem{Jazayeri:2022kjy}
S.~Jazayeri and S.~Renaux-Petel,
``Cosmological bootstrap in slow motion,''
JHEP \textbf{12}, 137 (2022)
doi:10.1007/JHEP12(2022)137
[arXiv:2205.10340 [hep-th]].

\bibitem{Jazayeri:2023xcj}
S.~Jazayeri, S.~Renaux-Petel and D.~Werth,
``Shapes of the cosmological low-speed collider,''
JCAP \textbf{12}, 035 (2023)
doi:10.1088/1475-7516/2023/12/035
[arXiv:2307.01751 [hep-th]].

\bibitem{Maldacena:2002vr}
J.~M.~Maldacena,
``Non-Gaussian features of primordial fluctuations in single field inflationary models,''
JHEP \textbf{05}, 013 (2003)
doi:10.1088/1126-6708/2003/05/013
[arXiv:astro-ph/0210603 [astro-ph]].

\bibitem{Arroja:2011yj}
F.~Arroja and T.~Tanaka,
``A note on the role of the boundary terms for the non-Gaussianity in general k-inflation,''
JCAP \textbf{05}, 005 (2011)
doi:10.1088/1475-7516/2011/05/005
[arXiv:1103.1102 [astro-ph.CO]].

\bibitem{Renaux-Petel:2011zgy}
S.~Renaux-Petel,
``On the redundancy of operators and the bispectrum in the most general second-order scalar-tensor theory,''
JCAP \textbf{02}, 020 (2012)
doi:10.1088/1475-7516/2012/02/020
[arXiv:1107.5020 [astro-ph.CO]].

\bibitem{Bjorkmo:2019fls}
T.~Bjorkmo,
``Rapid-Turn Inflationary Attractors,''
Phys. Rev. Lett. \textbf{122}, no.25, 251301 (2019)
doi:10.1103/PhysRevLett.122.251301
[arXiv:1902.10529 [hep-th]].

\bibitem{Garcia-Saenz:2018ifx}
S.~Garcia-Saenz, S.~Renaux-Petel and J.~Ronayne,
``Primordial fluctuations and non-Gaussianities in sidetracked inflation,''
JCAP \textbf{07}, 057 (2018)
doi:10.1088/1475-7516/2018/07/057
[arXiv:1804.11279 [astro-ph.CO]].

\bibitem{Brown:2017osf}
A.~R.~Brown,
``Hyperbolic Inflation,''
Phys. Rev. Lett. \textbf{121}, no.25, 251601 (2018)
doi:10.1103/PhysRevLett.121.251601
[arXiv:1705.03023 [hep-th]].

\bibitem{Fumagalli:2019noh}
J.~Fumagalli, S.~Garcia-Saenz, L.~Pinol, S.~Renaux-Petel and J.~Ronayne,
``Hyper-Non-Gaussianities in Inflation with Strongly Nongeodesic Motion,''
Phys. Rev. Lett. \textbf{123}, no.20, 201302 (2019)
doi:10.1103/PhysRevLett.123.201302
[arXiv:1902.03221 [hep-th]].

\bibitem{Garcia-Saenz:2018vqf}
S.~Garcia-Saenz and S.~Renaux-Petel,
``Flattened non-Gaussianities from the effective field theory of inflation with imaginary speed of sound,''
JCAP \textbf{11}, 005 (2018)
doi:10.1088/1475-7516/2018/11/005
[arXiv:1805.12563 [hep-th]].

\bibitem{Weinberg:2005vy}
S.~Weinberg,
``Quantum contributions to cosmological correlations,''
Phys. Rev. D \textbf{72}, 043514 (2005)
doi:10.1103/PhysRevD.72.043514
[arXiv:hep-th/0506236 [hep-th]].

\bibitem{Animali:2022lig}
C.~Animali, P.~Conzinu and G.~Marozzi,
``On adiabatic renormalization with a physically motivated infrared cut-off,''
JCAP \textbf{05}, no.05, 026 (2022)
doi:10.1088/1475-7516/2022/05/026
[arXiv:2201.05602 [gr-qc]].

\bibitem{Renaux-Petel:2015mga}
S.~Renaux-Petel and K.~Turzy\'nski,
``Geometrical Destabilization of Inflation,''
Phys. Rev. Lett. \textbf{117}, no.14, 141301 (2016)
doi:10.1103/PhysRevLett.117.141301
[arXiv:1510.01281 [astro-ph.CO]].

\bibitem{Planck:2018vyg}
N.~Aghanim \textit{et al.} [Planck],
``Planck 2018 results. VI. Cosmological parameters,''
Astron. Astrophys. \textbf{641}, A6 (2020)
[erratum: Astron. Astrophys. \textbf{652}, C4 (2021)]
doi:10.1051/0004-6361/201833910
[arXiv:1807.06209 [astro-ph.CO]].


\bibitem{Chen:2009we}
X.~Chen and Y.~Wang,
``Large non-Gaussianities with Intermediate Shapes from Quasi-Single Field Inflation,''
Phys. Rev. D \textbf{81}, 063511 (2010)
doi:10.1103/PhysRevD.81.063511
[arXiv:0909.0496 [astro-ph.CO]].

\bibitem{Baumann:2011nk}
D.~Baumann and D.~Green,
``Signatures of Supersymmetry from the Early Universe,''
Phys. Rev. D \textbf{85}, 103520 (2012)
doi:10.1103/PhysRevD.85.103520
[arXiv:1109.0292 [hep-th]].

\bibitem{Noumi:2012vr}
T.~Noumi, M.~Yamaguchi and D.~Yokoyama,
``Effective field theory approach to quasi-single field inflation and effects of heavy fields,''
JHEP \textbf{06}, 051 (2013)
doi:10.1007/JHEP06(2013)051
[arXiv:1211.1624 [hep-th]].

\bibitem{Arkani-Hamed:2015bza}
N.~Arkani-Hamed and J.~Maldacena,
``Cosmological Collider Physics,''
[arXiv:1503.08043 [hep-th]].

\bibitem{Chen:2009zp}
X.~Chen and Y.~Wang,
``Quasi-Single Field Inflation and Non-Gaussianities,''
JCAP \textbf{04}, 027 (2010)
doi:10.1088/1475-7516/2010/04/027
[arXiv:0911.3380 [hep-th]].

\bibitem{Assassi:2012zq}
V.~Assassi, D.~Baumann and D.~Green,
``On Soft Limits of Inflationary Correlation Functions,''
JCAP \textbf{11}, 047 (2012)
doi:10.1088/1475-7516/2012/11/047
[arXiv:1204.4207 [hep-th]].

\bibitem{Sefusatti:2012ye}
E.~Sefusatti, J.~R.~Fergusson, X.~Chen and E.~P.~S.~Shellard,
``Effects and Detectability of Quasi-Single Field Inflation in the Large-Scale Structure and Cosmic Microwave Background,''
JCAP \textbf{08}, 033 (2012)
doi:10.1088/1475-7516/2012/08/033
[arXiv:1204.6318 [astro-ph.CO]].

\bibitem{Norena:2012yi}
J.~Norena, L.~Verde, G.~Barenboim and C.~Bosch,  { {Prospects for
  constraining the shape of non-Gaussianity with the scale-dependent bias}},
  JCAP {\bf 08} (2012) 019 doi:10.1088/1475-7516/2012/08/019 [arXiv:1204.6324 [astro-ph.CO]]

\bibitem{Emami:2013lma}
R.~Emami,
``Spectroscopy of Masses and Couplings during Inflation,''
JCAP \textbf{04}, 031 (2014)
doi:10.1088/1475-7516/2014/04/031
[arXiv:1311.0184 [hep-th]].

\bibitem{Liu:2015tza}
J.~Liu, Y.~Wang and S.~Zhou,
``Inflation with Massive Vector Fields,''
JCAP \textbf{08}, 033 (2015)
doi:10.1088/1475-7516/2015/08/033
[arXiv:1502.05138 [hep-th]].

\bibitem{Dimastrogiovanni:2015pla}
E.~Dimastrogiovanni, M.~Fasiello and M.~Kamionkowski,
``Imprints of Massive Primordial Fields on Large-Scale Structure,''
JCAP \textbf{02}, 017 (2016)
doi:10.1088/1475-7516/2016/02/017
[arXiv:1504.05993 [astro-ph.CO]].

\bibitem{Schmidt:2015xka}
F.~Schmidt, N.~E.~Chisari and C.~Dvorkin,
``Imprint of inflation on galaxy shape correlations,''
JCAP \textbf{10}, 032 (2015)
doi:10.1088/1475-7516/2015/10/032
[arXiv:1506.02671 [astro-ph.CO]].

\bibitem{Chen:2015lza}
X.~Chen, M.~H.~Namjoo and Y.~Wang,
``Quantum Primordial Standard Clocks,''
JCAP \textbf{02}, 013 (2016)
doi:10.1088/1475-7516/2016/02/013
[arXiv:1509.03930 [astro-ph.CO]].

\bibitem{Bonga:2015urq}
B.~Bonga, S.~Brahma, A.~S.~Deutsch and S.~Shandera,
``Cosmic variance in inflation with two light scalars,''
JCAP \textbf{05}, 018 (2016)
doi:10.1088/1475-7516/2016/05/018
[arXiv:1512.05365 [astro-ph.CO]].

\bibitem{Delacretaz:2015edn}
L.~V.~Delacretaz, T.~Noumi and L.~Senatore,
``Boost Breaking in the EFT of Inflation,''
JCAP \textbf{02}, 034 (2017)
doi:10.1088/1475-7516/2017/02/034
[arXiv:1512.04100 [hep-th]].

\bibitem{Flauger:2016idt}
R.~Flauger, M.~Mirbabayi, L.~Senatore and E.~Silverstein,
``Productive Interactions: heavy particles and non-Gaussianity,''
JCAP \textbf{10}, 058 (2017)
doi:10.1088/1475-7516/2017/10/058
[arXiv:1606.00513 [hep-th]].

\bibitem{Lee:2016vti}
H.~Lee, D.~Baumann and G.~L.~Pimentel,
``Non-Gaussianity as a Particle Detector,''
JHEP \textbf{12}, 040 (2016)
doi:10.1007/JHEP12(2016)040
[arXiv:1607.03735 [hep-th]].

\bibitem{Delacretaz:2016nhw}
L.~V.~Delacretaz, V.~Gorbenko and L.~Senatore,
``The Supersymmetric Effective Field Theory of Inflation,''
JHEP \textbf{03}, 063 (2017)
doi:10.1007/JHEP03(2017)063
[arXiv:1610.04227 [hep-th]].

\bibitem{Meerburg:2016zdz}
P.~D.~Meerburg, M.~M\"unchmeyer, J.~B.~Mu\~noz and X.~Chen,
``Prospects for Cosmological Collider Physics,''
JCAP \textbf{03}, 050 (2017)
doi:10.1088/1475-7516/2017/03/050
[arXiv:1610.06559 [astro-ph.CO]].

\bibitem{Chen:2016uwp}
X.~Chen, Y.~Wang and Z.~Z.~Xianyu,
``Standard Model Background of the Cosmological Collider,''
Phys. Rev. Lett. \textbf{118}, no.26, 261302 (2017)
doi:10.1103/PhysRevLett.118.261302
[arXiv:1610.06597 [hep-th]].

\bibitem{Chen:2016hrz}
X.~Chen, Y.~Wang and Z.~Z.~Xianyu,
``Standard Model Mass Spectrum in Inflationary Universe,''
JHEP \textbf{04}, 058 (2017)
doi:10.1007/JHEP04(2017)058
[arXiv:1612.08122 [hep-th]].

\bibitem{An:2017hlx}
H.~An, M.~McAneny, A.~K.~Ridgway and M.~B.~Wise,
``Quasi Single Field Inflation in the non-perturbative regime,''
JHEP \textbf{06}, 105 (2018)
doi:10.1007/JHEP06(2018)105
[arXiv:1706.09971 [hep-ph]].

\bibitem{Tong:2017iat}
X.~Tong, Y.~Wang and S.~Zhou,
``On the Effective Field Theory for Quasi-Single Field Inflation,''
JCAP \textbf{11}, 045 (2017)
doi:10.1088/1475-7516/2017/11/045
[arXiv:1708.01709 [astro-ph.CO]].

\bibitem{Iyer:2017qzw}
A.~V.~Iyer, S.~Pi, Y.~Wang, Z.~Wang and S.~Zhou,
``Strongly Coupled Quasi-Single Field Inflation,''
JCAP \textbf{01}, 041 (2018)
doi:10.1088/1475-7516/2018/01/041
[arXiv:1710.03054 [hep-th]].

\bibitem{An:2017rwo}
H.~An, M.~McAneny, A.~K.~Ridgway and M.~B.~Wise,
``Non-Gaussian Enhancements of Galactic Halo Correlations in Quasi-Single Field Inflation,''
Phys. Rev. D \textbf{97}, no.12, 123528 (2018)
doi:10.1103/PhysRevD.97.123528
[arXiv:1711.02667 [hep-ph]].

\bibitem{Kumar:2017ecc}
S.~Kumar and R.~Sundrum,
``Heavy-Lifting of Gauge Theories By Cosmic Inflation,''
JHEP \textbf{05}, 011 (2018)
doi:10.1007/JHEP05(2018)011
[arXiv:1711.03988 [hep-ph]].

\bibitem{Riquelme:2017bxt}
S.~Riquelme M.,
``Non-Gaussianities in a two-field generalization of Natural Inflation,''
JCAP \textbf{04}, 027 (2018)
doi:10.1088/1475-7516/2018/04/027
[arXiv:1711.08549 [astro-ph.CO]].

\bibitem{Saito:2018omt}
R.~Saito and T.~Kubota,
``Heavy Particle Signatures in Cosmological Correlation Functions with Tensor Modes,''
JCAP \textbf{06}, 009 (2018)
doi:10.1088/1475-7516/2018/06/009
[arXiv:1804.06974 [hep-th]].

\bibitem{Cabass:2018roz}
G.~Cabass, E.~Pajer and F.~Schmidt,
``Imprints of Oscillatory Bispectra on Galaxy Clustering,''
JCAP \textbf{09}, 003 (2018)
doi:10.1088/1475-7516/2018/09/003
[arXiv:1804.07295 [astro-ph.CO]].

\bibitem{Dimastrogiovanni:2018uqy}
E.~Dimastrogiovanni, M.~Fasiello and G.~Tasinato,
``Probing the inflationary particle content: extra spin-2 field,''
JCAP \textbf{08}, 016 (2018)
doi:10.1088/1475-7516/2018/08/016
[arXiv:1806.00850 [astro-ph.CO]].

\bibitem{Bordin:2018pca}
L.~Bordin, P.~Creminelli, A.~Khmelnitsky and L.~Senatore,
``Light Particles with Spin in Inflation,''
JCAP \textbf{10}, 013 (2018)
doi:10.1088/1475-7516/2018/10/013
[arXiv:1806.10587 [hep-th]].

\bibitem{Arkani-Hamed:2018kmz}
N.~Arkani-Hamed, D.~Baumann, H.~Lee and G.~L.~Pimentel,
``The Cosmological Bootstrap: Inflationary Correlators from Symmetries and Singularities,''
JHEP \textbf{04}, 105 (2020)
doi:10.1007/JHEP04(2020)105
[arXiv:1811.00024 [hep-th]].

\bibitem{Kumar:2018jxz}
S.~Kumar and R.~Sundrum,
``Seeing Higher-Dimensional Grand Unification In Primordial Non-Gaussianities,''
JHEP \textbf{04}, 120 (2019)
doi:10.1007/JHEP04(2019)120
[arXiv:1811.11200 [hep-ph]].

\bibitem{Goon:2018fyu}
G.~Goon, K.~Hinterbichler, A.~Joyce and M.~Trodden,
``Shapes of gravity: Tensor non-Gaussianity and massive spin-2 fields,''
JHEP \textbf{10}, 182 (2019)
doi:10.1007/JHEP10(2019)182
[arXiv:1812.07571 [hep-th]].

\bibitem{Wu:2018lmx}
Y.~P.~Wu,
``Higgs as heavy-lifted physics during inflation,''
JHEP \textbf{04}, 125 (2019)
doi:10.1007/JHEP04(2019)125
[arXiv:1812.10654 [hep-ph]].

\bibitem{Chua:2018dqh}
W.~Z.~Chua, Q.~Ding, Y.~Wang and S.~Zhou,
``Imprints of Schwinger Effect on Primordial Spectra,''
JHEP \textbf{04}, 066 (2019)
doi:10.1007/JHEP04(2019)066
[arXiv:1810.09815 [hep-th]].

\bibitem{Wang:2018tbf}
Y.~Wang, Y.~P.~Wu, J.~Yokoyama and S.~Zhou,
``Hybrid Quasi-Single Field Inflation,''
JCAP \textbf{07}, 068 (2018)
doi:10.1088/1475-7516/2018/07/068
[arXiv:1804.07541 [astro-ph.CO]].

\bibitem{McAneny:2019epy}
M.~McAneny and A.~K.~Ridgway,
%``New Shapes of Primordial Non-Gaussianity from Quasi-Single Field Inflation with Multiple Isocurvatons,''
Phys. Rev. D \textbf{100}, no.4, 043534 (2019)
doi:10.1103/PhysRevD.100.043534
[arXiv:1903.11607 [astro-ph.CO]].

\bibitem{Li:2019ves}
L.~Li, T.~Nakama, C.~M.~Sou, Y.~Wang and S.~Zhou,
``Gravitational Production of Superheavy Dark Matter and Associated Cosmological Signatures,''
JHEP \textbf{07}, 067 (2019)
doi:10.1007/JHEP07(2019)067
[arXiv:1903.08842 [astro-ph.CO]].

\bibitem{Kim:2019wjo}
S.~Kim, T.~Noumi, K.~Takeuchi and S.~Zhou,
``Heavy Spinning Particles from Signs of Primordial Non-Gaussianities: Beyond the Positivity Bounds,''
JHEP \textbf{12}, 107 (2019)
doi:10.1007/JHEP12(2019)107
[arXiv:1906.11840 [hep-th]].

\bibitem{Sleight:2019mgd}
C.~Sleight,
``A Mellin Space Approach to Cosmological Correlators,''
JHEP \textbf{01}, 090 (2020)
doi:10.1007/JHEP01(2020)090
[arXiv:1906.12302 [hep-th]].

\bibitem{Biagetti:2019bnp}
M.~Biagetti,
``The Hunt for Primordial Interactions in the Large Scale Structures of the Universe,''
Galaxies \textbf{7}, no.3, 71 (2019)
doi:10.3390/galaxies7030071
[arXiv:1906.12244 [astro-ph.CO]].

\bibitem{Sleight:2019hfp}
C.~Sleight and M.~Taronna,
``Bootstrapping Inflationary Correlators in Mellin Space,''
JHEP \textbf{02}, 098 (2020)
doi:10.1007/JHEP02(2020)098
[arXiv:1907.01143 [hep-th]].

\bibitem{Welling:2019bib}
Y.~Welling,
``Simple, exact model of quasisingle field inflation,''
Phys. Rev. D \textbf{101}, no.6, 063535 (2020)
doi:10.1103/PhysRevD.101.063535
[arXiv:1907.02951 [astro-ph.CO]].

\bibitem{Alexander:2019vtb}
S.~Alexander, S.~J.~Gates, L.~Jenks, K.~Koutrolikos and E.~McDonough,
``Higher Spin Supersymmetry at the Cosmological Collider: Sculpting SUSY Rilles in the CMB,''
JHEP \textbf{10}, 156 (2019)
doi:10.1007/JHEP10(2019)156
[arXiv:1907.05829 [hep-th]].

\bibitem{Lu:2019tjj}
S.~Lu, Y.~Wang and Z.~Z.~Xianyu,
``A Cosmological Higgs Collider,''
JHEP \textbf{02}, 011 (2020)
doi:10.1007/JHEP02(2020)011
[arXiv:1907.07390 [hep-th]].

\bibitem{Hook:2019zxa}
A.~Hook, J.~Huang and D.~Racco,
``Searches for other vacua. Part II. A new Higgstory at the cosmological collider,''
JHEP \textbf{01}, 105 (2020)
doi:10.1007/JHEP01(2020)105
[arXiv:1907.10624 [hep-ph]].

\bibitem{Hook:2019vcn}
A.~Hook, J.~Huang and D.~Racco,
``Minimal signatures of the Standard Model in non-Gaussianities,''
Phys. Rev. D \textbf{101}, no.2, 023519 (2020)
doi:10.1103/PhysRevD.101.023519
[arXiv:1908.00019 [hep-ph]].

\bibitem{ScheihingHitschfeld:2019tzr}
B.~Scheihing Hitschfeld,
``Revealing the Structure of the Inflationary Landscape through Primordial non-Gaussianity,''
[arXiv:1909.11223 [astro-ph.CO]].

\bibitem{Baumann:2019oyu}
D.~Baumann, C.~Duaso Pueyo, A.~Joyce, H.~Lee and G.~L.~Pimentel,
``The cosmological bootstrap: weight-shifting operators and scalar seeds,''
JHEP \textbf{12}, 204 (2020)
doi:10.1007/JHEP12(2020)204
[arXiv:1910.14051 [hep-th]].

\bibitem{Wang:2019gbi}
L.~T.~Wang and Z.~Z.~Xianyu,
``In Search of Large Signals at the Cosmological Collider,''
JHEP \textbf{02}, 044 (2020)
doi:10.1007/JHEP02(2020)044
[arXiv:1910.12876 [hep-ph]].

\bibitem{Liu:2019fag}
T.~Liu, X.~Tong, Y.~Wang and Z.~Z.~Xianyu,
``Probing P and CP Violations on the Cosmological Collider,''
JHEP \textbf{04}, 189 (2020)
doi:10.1007/JHEP04(2020)189
[arXiv:1909.01819 [hep-ph]].

\bibitem{Wang:2019gok}
D.~G.~Wang,
``On the inflationary massive field with a curved field manifold,''
JCAP \textbf{01}, 046 (2020)
doi:10.1088/1475-7516/2020/01/046
[arXiv:1911.04459 [astro-ph.CO]].

\bibitem{Wang:2020uic}
Y.~Wang and Y.~Zhu,
``Cosmological Collider Signatures of Massive Vectors from Non-Gaussian Gravitational Waves,''
JCAP \textbf{04}, 049 (2020)
doi:10.1088/1475-7516/2020/04/049
[arXiv:2001.03879 [astro-ph.CO]].

\bibitem{Jazayeri:2023kji}
S.~Jazayeri, S.~Renaux-Petel, X.~Tong, D.~Werth and Y.~Zhu,
v``Parity Violation from Emergent Non-Locality During Inflation,''
[arXiv:2308.11315 [hep-th]].

\bibitem{Tong:2021wai}
X.~Tong, Y.~Wang and Y.~Zhu,
``Cutting rule for cosmological collider signals: a bulk evolution perspective,''
JHEP \textbf{03}, 181 (2022)
doi:10.1007/JHEP03(2022)181
[arXiv:2112.03448 [hep-th]].

\bibitem{Pimentel:2022fsc}
G.~L.~Pimentel and D.~G.~Wang,
``Boostless cosmological collider bootstrap,''
JHEP \textbf{10}, 177 (2022)
doi:10.1007/JHEP10(2022)177
[arXiv:2205.00013 [hep-th]].

\bibitem{Wang:2022eop}
D.~G.~Wang, G.~L.~Pimentel and A.~Ach\'ucarro,
``Bootstrapping multi-field inflation: non-Gaussianities from light scalars revisited,''
JCAP \textbf{05}, 043 (2023)
doi:10.1088/1475-7516/2023/05/043
[arXiv:2212.14035 [astro-ph.CO]].

\bibitem{Werth:2023pfl}
D.~Werth, L.~Pinol and S.~Renaux-Petel,
``Cosmological Flow of Primordial Correlators,''
[arXiv:2302.00655 [hep-th]].

\bibitem{Niu:2022quw}
X.~Niu, M.~H.~Rahat, K.~Srinivasan and W.~Xue,
``Gravitational wave probes of massive gauge bosons at the cosmological collider,''
JCAP \textbf{02}, 013 (2023)
doi:10.1088/1475-7516/2023/02/013
[arXiv:2211.14331 [hep-ph]].

\bibitem{Niu:2022fki}
X.~Niu, M.~H.~Rahat, K.~Srinivasan and W.~Xue,
``Parity-odd and even trispectrum from axion inflation,''
JCAP \textbf{05}, 018 (2023)
doi:10.1088/1475-7516/2023/05/018
[arXiv:2211.14324 [hep-ph]].

\bibitem{Deskins:2013dwa}
J.~T.~Deskins, J.~T.~Giblin and R.~R.~Caldwell,
``Gauge Field Preheating at the End of Inflation,''
Phys. Rev. D \textbf{88}, no.6, 063530 (2013)
doi:10.1103/PhysRevD.88.063530
[arXiv:1305.7226 [astro-ph.CO]].

\bibitem{Domcke:2020zez}
V.~Domcke, V.~Guidetti, Y.~Welling and A.~Westphal,
``Resonant backreaction in axion inflation,''
JCAP \textbf{09}, 009 (2020)
doi:10.1088/1475-7516/2020/09/009
[arXiv:2002.02952 [astro-ph.CO]].

\bibitem{Caravano:2022epk}
A.~Caravano, E.~Komatsu, K.~D.~Lozanov and J.~Weller,
``Lattice Simulations of Axion-U(1) Inflation,''
[arXiv:2204.12874 [astro-ph.CO]].

\bibitem{Figueroa:2023oxc}
D.~G.~Figueroa, J.~Lizarraga, A.~Urio and J.~Urrestilla,
``The strong backreaction regime in axion inflation,''
[arXiv:2303.17436 [astro-ph.CO]].

\bibitem{Ferreira:2015omg}
R.~Z.~Ferreira, J.~Ganc, J.~Nore\~na and M.~S.~Sloth,
``On the validity of the perturbative description of axions during inflation,''
JCAP \textbf{04}, 039 (2016)
[erratum: JCAP \textbf{10}, E01 (2016)]
doi:10.1088/1475-7516/2016/04/039
[arXiv:1512.06116 [astro-ph.CO]].

\bibitem{Peloso:2016gqs}
M.~Peloso, L.~Sorbo and C.~Unal,
``Rolling axions during inflation: perturbativity and signatures,''
JCAP \textbf{09}, 001 (2016)
doi:10.1088/1475-7516/2016/09/001
[arXiv:1606.00459 [astro-ph.CO]].






% Please avoid comments such as "For a review'', "For some examples",
% "and references therein" or move them in the text. In general,
% please leave only references in the bibliography and move all
% accessory text in footnotes.

% Also, please have only one work for each \bibitem.


\end{thebibliography}
\end{document}